\documentclass[twocolumn]{aastex62}

\newcommand{\thisgrb}{GRB~190114C~}
\newcommand{\oldgrb}{GRB~160509A~}

\newcommand{\fermi}{{\em Fermi~}}
\newcommand{\swift}{{\em Swift~}}
\newcommand{\sw}[1]{\texttt{#1}}
\graphicspath{{./}{figures/}}

\shorttitle{MAGICal \thisgrb prompt emission spectrum}
\shortauthors{Vikas Chand et al.}

\begin{document}

\title{MAGICal GRB 190114C: Implications of cutoff in the spectrum at sub-$GeV$ energies}

\author[0000-0002-7876-7362]{Vikas Chand$^\dagger$}
\affiliation{School of Physics \& Astronomy, Sun Yat Sen University, Guangzhou 510275, China}
\affiliation{Tata Institute of Fundamental Research, Mumbai, India}
\author[0000-0001-8922-8391]{Partha Sarathi Pal}
\affiliation{School of Physics \& Astronomy, Sun Yat Sen University, Guangzhou 510275, China}
\author[0000-0002-4371-2417]{Ankush Banerjee}
\noaffiliation{}
\author{Vidushi Sharma$^{*}$}
\affiliation{Inter University Centre for Astronomy and Astrophysics, Pune, India}
\author[0000-0002-1262-7375]{P.~H.~T.~Tam$^{!}$}
\affiliation{School of Physics \& Astronomy, Sun Yat Sen University, Guangzhou 510275, China}
\author{Xinbo He}
\affiliation{School of Physics \& Astronomy, Sun Yat Sen University, Guangzhou 510275, China}

\email{$^\dagger$vikas.chand.physics@gmail.com}
\email{$^*$vidushi@iucaa.in}
\email{$^!$tanbxuan@mail.sysu.edu.cn}

\begin{abstract}
\thisgrb  is an unusual gamma-ray burst (GRB) due to its detection at sub-$TeV$ energies by MAGIC, seen at redshift z = 0.42. This burst is one of the brightest GRB detected by \fermi. 
A joint GBM-LAT analysis of the prompt emission reveals the presence of sub-$GeV$ spectral cutoff when the LAT \emph{low-energy events} (LLE) data is also examined. A similar high-energy cutoff was likewise reported in \oldgrb and GRB 100724B earlier, as well as handful of other sources. The cutoff can be explained by the intrinsic opacity due to pair production within the emitting region. \thisgrb shows a transition from non-thermal to a quasi-thermal-like spectrum and a radiation component that can be attributed to afterglow. Based on spectral analysis, we constrain the site of the prompt emission and $Lorentz$ factor.
Knowing that sub-$TeV$ photons are detected in MAGIC, we perceive that the observed spectrum is indeed an overlap from two emission sites, where the emission observed in \fermi is more consistent with prompt emission produced via photospheric dissipation along with a concurrent component from the external shock.
\end{abstract}


\section{Introduction}
\label{sec:intro}
Gamma-ray burst (GRB) prompt emission spectra are traditionally modeled by the Band function \citep{Band:1993}. 
However, deviations from the Band function were observed 
and reported previously, such as the presence of an extra thermal component \citep{Ryde:2005, Page:2011, Guiriec:2011}, spectral breaks \citep{Oganesyan:2017, Oganesyan:2017a}, a Band function with high-energy cutoff \citep{Ackermann:2013, Tang:2015ApJ, Vianello:2017}, and 
multiple thermal $+$ non-thermal components \citep{Guiriec:2015ApJ, Guiriec:2015b, Guiriec:2016ApJ, BR:2015}. Observational evidence, however, could be 
affected by selection effects. Two very different models are sometimes barely distinguishable when folded with the response of a detector (see, e.g., \citealt{ZhangBB:2016ApJ, Vianello:2017, Burgess:2019A&A629A}). 
Time-resolved spectral analysis, though a very reliable method to understand the emission mechanism, has implementation limits because of poor statistics. A break at low energy is seen in GRBs observed simultaneously in soft and hard X-rays in \swift and at low and high-energy bands in \fermi \citep{Oganesyan:2017ApJ, Ravasio2019A&A}. However, multiple components in this range are also observed \citep{Basak:2013ApJ, Iyyani:2013}. The presence of a thermal component and its effect on the non-thermal spectral emission have also been studied \citep[e.g.,][]{Li:2019arXiv, Guiriec:2011, Axelsson:2012ApJ, Guiriec:2013}, and the thermal components are in general considered to be signatures of photospheric emission. 

On the theoretical side, consensus is building over two contending models, photospheric emission (dissipative or non-dissipative) and synchrotron emission in many possible settings (see, \citealt{Kumar:2015PhR} for a review).
Recently, a synchrotron model from a cooling population of electrons was used to study the prompt emission of the GRBs. Most of the GRB spectra used in these studies are consistent with synchrotron cooling \citep{Burgess:2018arXiv, Oganesyan:2019A&A}. Photospheric models are also used in some studies and can well explain the data well \citep{Vurm:2016, Vianello:2017, Ahlgren:2019ApJ, Zeynep:2020arXiv}.

The sub-$GeV$ radiation from the bright burst with a good count statistics can give important insights.
The question of whether the $\sim100~MeV$ emission has an external shock origin or internal dissipation origin is still under debate \citep[see][and references therein]{Tang:2015ApJ}. Time-dependent broadband model fits from $keV$ to $GeV$ energies can help us to distinguish these two origins. The spectral cutoffs, if observed, can also play a pivotal role in pinning down the site of the prompt emission production in GRBs \citep{Gupta:2008MNRAS}. The spectral cutoffs can also be decisively important in understanding 
the origin of the radiation and components of a GRB jet \citep{Zhang:2009ApJ}.
Additionally, in the cases where both the external and internal dissipation is contributing to the $\sim100~MeV$ emission, model fits with short time slices (e.g., 1 $s$) can reveal the time evolution of both contributions as well as the transition from prompt emission to the afterglow emission phase.

With the \fermi space observatory, the broadband spectrum of the GRBs can be studied from a few $keV$ up to hundreds of $GeV$s. 
Some bright bursts have shown bright emission in \fermi-LAT also, thus allowing enough photon statistics even in short time slices. GRB 160625B is one such example where the emission was seen in the sub-$GeV$ LAT band. A joint analysis of the emission observed in the \fermi detectors shows a cutoff in the spectra in the $\sim100~MeV$ energy range \citep{Wang:2017}. Similarly, \oldgrb is yet another example. A break similar to GRB 160625B exists for this GRB and GRB 100724B \citep{Vianello:2017}. 
A detailed analysis with LAT emission during the prompt emission can reveal that the contamination of the spectrum by lower energy components can lead to a dramatic evolution (discussed in Appendix \ref{sec:afterglow_transition}).
\thisgrb shows one bright pulse in the LAT.
Several spectral studies of \thisgrb have been performed (\citealt{Wang:2019arXiv}, \citealt{Ravasio:2019arXiv}), but caveats of these works include: (a) 
separate analysis instead of joint analysis and (b) wider time bins being chosen while performing the spectral analysis. There are hints from these studies though that the initial part of the LAT spectrum could be affected by the prompt emission spectrum.  

\thisgrb, in its multiwavelength spectral energy density, showed evidence for a double-peaked distribution with the first peak being the synchrotron emission. The second peak shows a very high-energy (VHE) emission in TeV energies and is explained by the synchrotron self-Comptonization process, theoretically predicted in a standard afterglow model \citep{MAGIC2019Natur_IC}. We present the evolution of the spectrum of the prompt emission by using joint GBM-LAT analysis, by trying several typical empirical spectral models. The observed sub-GeV cutoffs and the thermal components in the spectra are utilized to evaluate $Lorentz$ factor ($\Gamma$) in various scenarios and also to constrain the site of the prompt emission \citep{Lithwick:2001, Asaf:2007ApJ, Zhang:2009ApJ}. 

We summarize the major observations of \thisgrb in Section \ref{sec:observations}. We draw a parallel of \thisgrb with \oldgrb and also demonstrate the transition of the LAT emission from prompt to afterglow for the later in Section \ref{sec:analysis}. Results are presented and discussed in Sections \ref{sec:conclusions} and \ref{sec:discussion}, respectively. We also underline a systematic approach in Appendix \ref{sec:afterglow_transition} that can be followed to uncover a sub-GeV cutoff in the case of a simultaneous presence of external components.

\section{Observations}
\label{sec:observations}
\thisgrb triggered the \emph{Neil Gehrels Swift Observatory} - Burst Alert Telescope (BAT) at 20:57:03 UT ($T0$, trigger time), on 2019 January 14. Later, the optical counterpart was detected by several observatories in various bands, and with detection of absorption lines the redshift was found to be z = $0.4245\pm0.0005$  \citep{Castro-Tirado:2019GCN}. Surprisingly, MAGIC detected \thisgrb in the sub-$TeV$ energy domain starting at $T0$ $+$ 50 $s$. A clear excess of gamma ray events was detected with a significance of more than 20 $\sigma$ within the first 20 minutes with energies of the photons greater than $300$ $GeV$ \citep{Mirzoyan:2019GCN}. The \fermi GBM light curve shows a bright, multipeaked pulses from $T0$ $+$ 0 $s$ to $T0$ $+$ 15 $s$ followed by a fainter emission lasting up to $T0$ $+$ 200 $s$. The calculated T90 \citep{KoshutT90:1995GRB} duration of the light curve was found to be 116 $s$ ( within $50~-~300$ $KeV$), along with an energy fluence (within $10~-~1000$ $keV$) of (3.99$\times$10$^{-4}$ $\pm$ 8.10$\times$10$^{-7}$) $erg$ $cm^{-2}$ and the estimated isotropic energy release was 3$\times$10$^{53}$ $erg$. This source was also detected by AGILE/MCAL in the $0.4~-~100$ $MeV$ energy band for a duration of $6.2$ $s$ \citep{Ursi:2019GCN}. As observed by \emph{Konus-Wind}, the main burst showed a hard-spectrum multipeaked pulse starting from $T0$ to $T0$ + 6 $s$ with a fluence of ($4.83$ $\pm$ $0.10$) $\times$ 10$^{-4}$ $erg$ $cm^{-2}$ \citep{Frederiks:2019GCN}.

\section{Method and analysis}
\label{sec:analysis}
The data are obtained from the publicly available data archive on the \fermi \emph{Science Support Center} (FSSC) {\footnote{\url{https://fermi.gsfc.nasa.gov/ssc/}}}.
The spectra were reduced using \emph{Fermi science tools} software \sw{gtburst} by standard methodology\footnote{\url{https://fermi.gsfc.nasa.gov/ssc/data/analysis/scitools/gtburst.html}}.
For LAT, transient event class and its instrument response function \sw{P8\_TRANSIENT020} were used.
The spectral analysis is performed in \sw{XSPEC} \citep{Arnaud:1996}, and \sw{pgstat} was used for testing various models since the data is Poissonian and background is Gaussian background as it is derived from modeling the data in off-source intervals by a polynomial \footnote{\url{https://heasarc.gsfc.nasa.gov/xanadu/xspec/manual/node293.html}}. Furthermore, to fit the different components of the spectrum we used the Band model \citep{Band:1993} (B) for one, and B + power law with a multiplicative cutoff component (B + CPL) for the other. A model with exponential cutoff applied to the Band model (BC) is also used (see section \ref{sec:models} for the forms of the functions used. To find which model fits the data best, we used the Bayesian Information Criterion (BIC) and the Akaike Information Criterion (AIC).  
Given their properties, AIC is preferred to compare nonnested models such as Band function, or power law. Whereas, BIC is preferred when nested models such as blackbody added to a band function are compared \citep{Kass:1995}. 
The change in AIC or BIC can predict the model with strong correlation to the data. All errors are presented within 1 $\sigma$ (68$\%$ nominal) confidence levels.

We take brighter GBM-NaI detectors with off-axis angles less than $50^\circ$ and GBM-BGO covering the same hemisphere of spacecraft as NaI detectors. In case of \thisgrb, we have considered NaI 3, 4, 7, 8 and BGO 0, 1 (n3, n4, n7, n8, b0 and b1). LAT data is also used (both LLE and $>$ 100 $MeV$ data). Here, we would be referring to energies $>$ 100 $MeV$ as LAT-HE. To account for inter-instrument calibration, we applied a multiplicative constant factor (effective area correction factor) w.r.t the detector having the highest count rate. The factor is allowed to vary up to $\sim20$ - $30\%$ as the EAC constant factor is not expected to differ by more than $30\%$.


\subsection{\thisgrb}

\subsubsection{Joint GBM-LAT analysis}
\begin{figure*}
    \centering
    \includegraphics[scale=0.58]{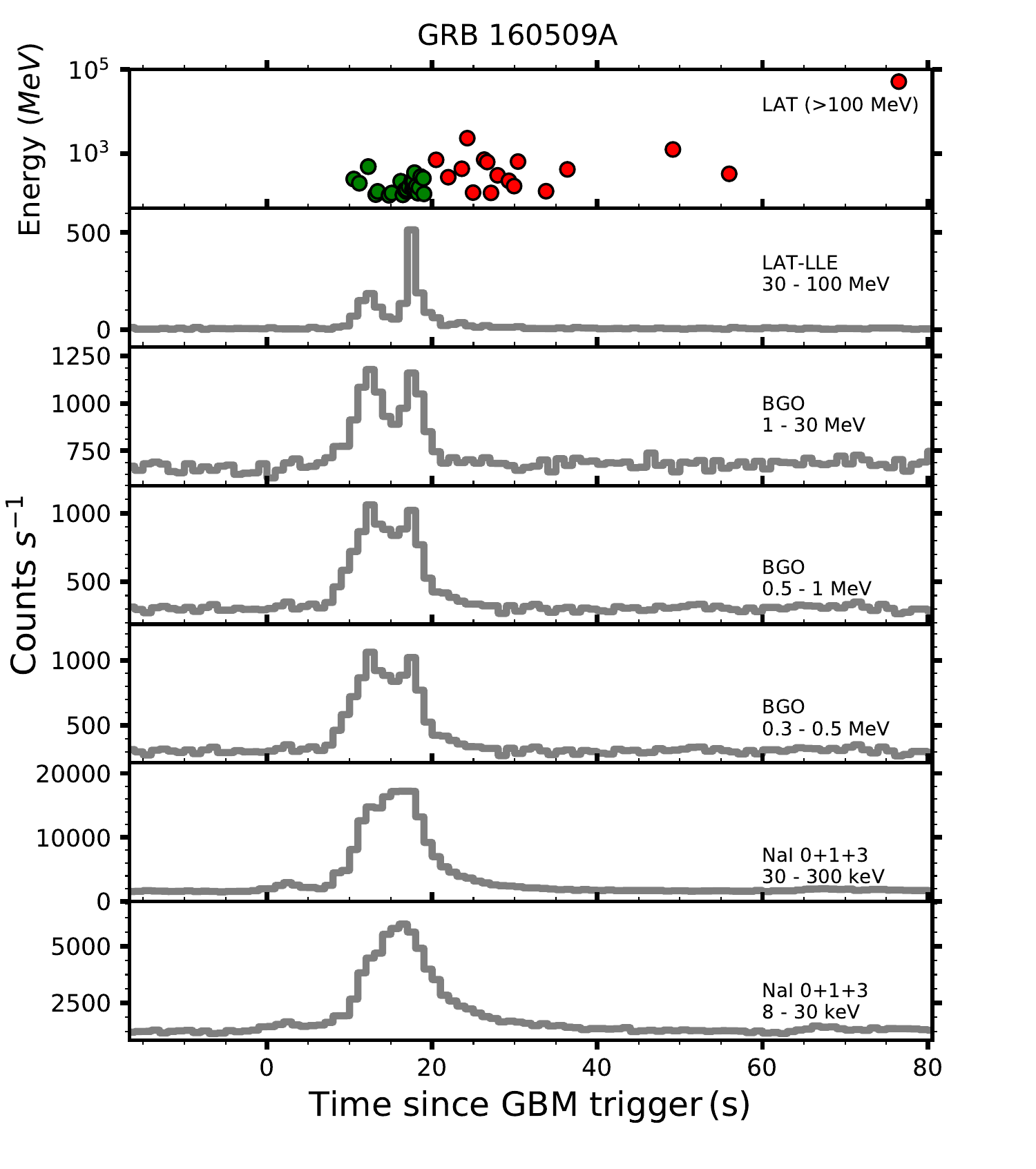}
    \includegraphics[scale=0.58]{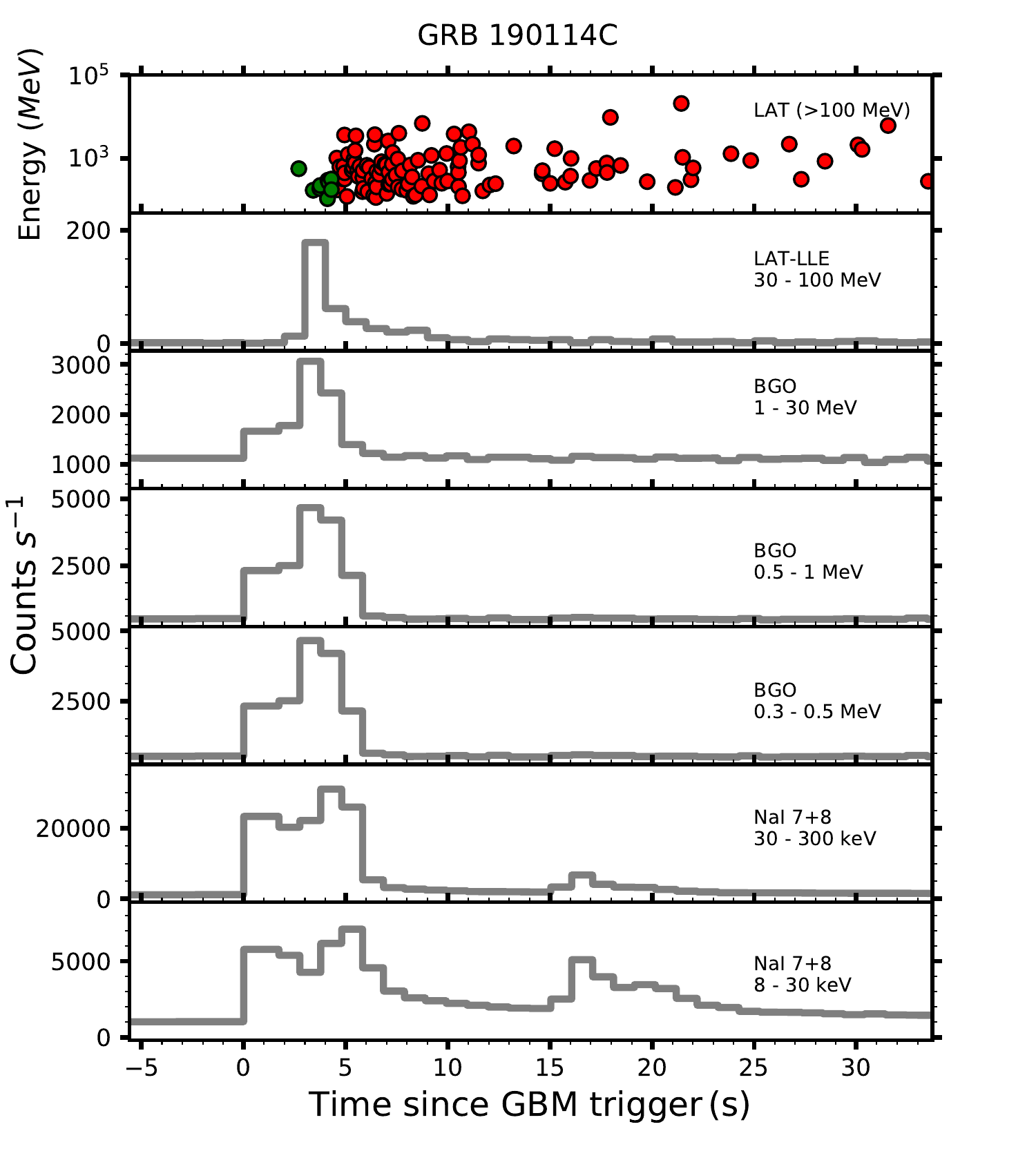}
    \caption{Light-curve evolution of \oldgrb and \thisgrb. In the case of \oldgrb the green filled circles are photons observed before 20 $s$, and 4.5 $s$ in the case of \thisgrb with $>90\%$ probability of their association with the GRBs respectively. Similarly, red filled circles are after these times and with $>90\%$ probability of their association with the GRB.}
    \label{fig:LCs}
\end{figure*}

The energy range 8 - 900 $keV$ was used for NaI detectors, $\sim$ 0.2 - 38 $MeV$ was used for the BGO detectors, 20 - 100 $MeV$ was used from LLE and $>$100 $MeV$ was used for LAT-HE.  
We neglected $\sim$ 30 - 40 $keV$ from our spectral analysis to exclude the 33.17 $keV$ K-edge feature. In Fig.~\ref{fig:LCs}, we can observe that contrary to \oldgrb, the initial emission in \thisgrb is limited to the 1 - 30 $MeV$ band only, however, the bright pulse during the peak finds its correspondence in the 30 - 100 $MeV$ LLE band, and in LAT-HE only some photons with relatively low energies are observed.
As in the case of \oldgrb, the later emission can contaminate the prompt emission (See Appendix \ref{sec:afterglow_transition} for more details). The presence of the afterglow component could be felt prominently at low energies during the prompt emission. 
This component noticeably pollutes the prompt spectrum after $4.8$ $s$ \citep{Ravasio:2019arXiv}. Interestingly, the LAT photon index also shows soft to hard evolution \citep{Wang:2019arXiv} similar to \oldgrb. We thus explore the joint GBM-LAT data for the possibility of a spectral cutoff in the prompt emission. 

\begin{figure*}
\centering
\includegraphics[scale=0.325, angle=270]{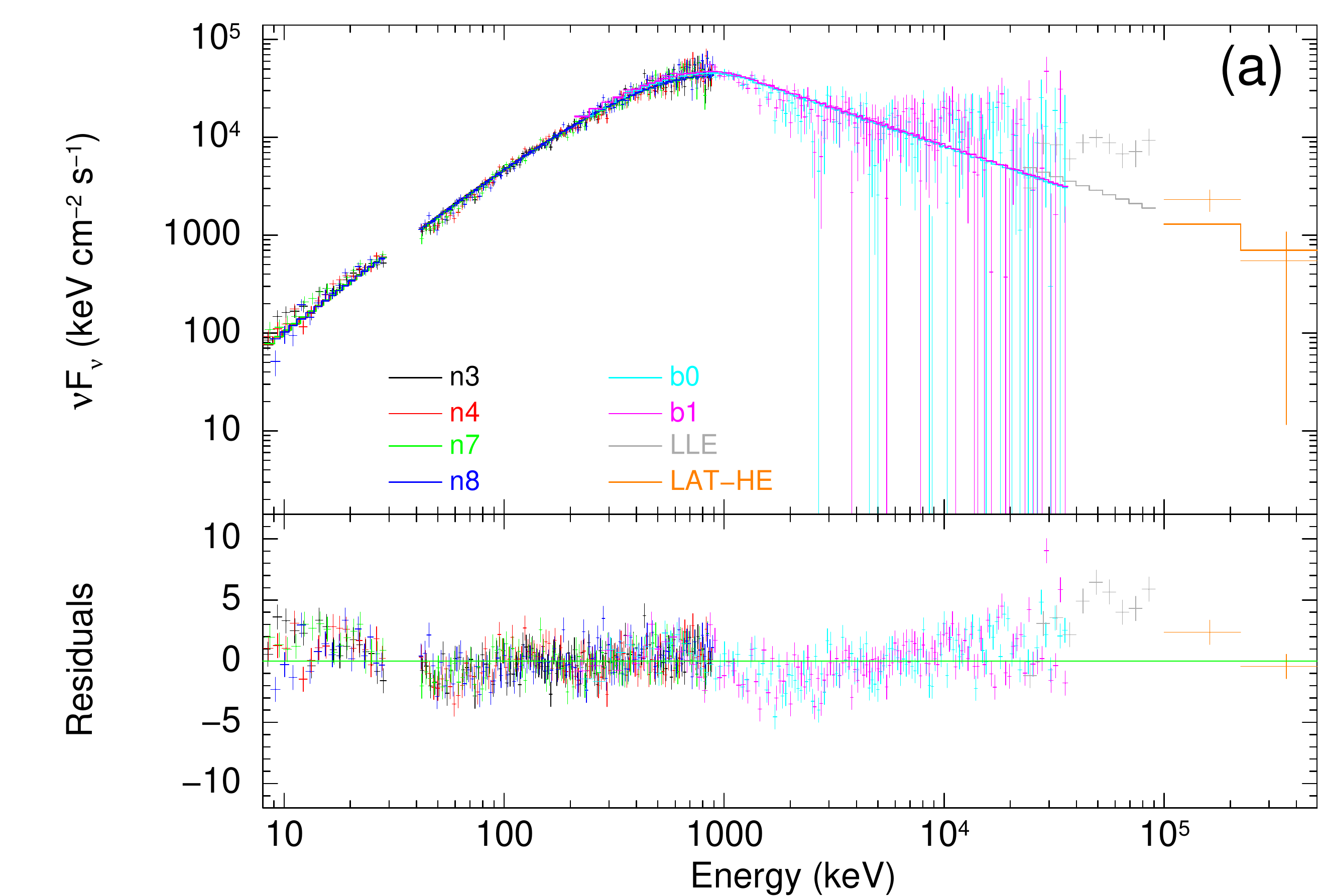}
\includegraphics[scale=0.325, angle=270]{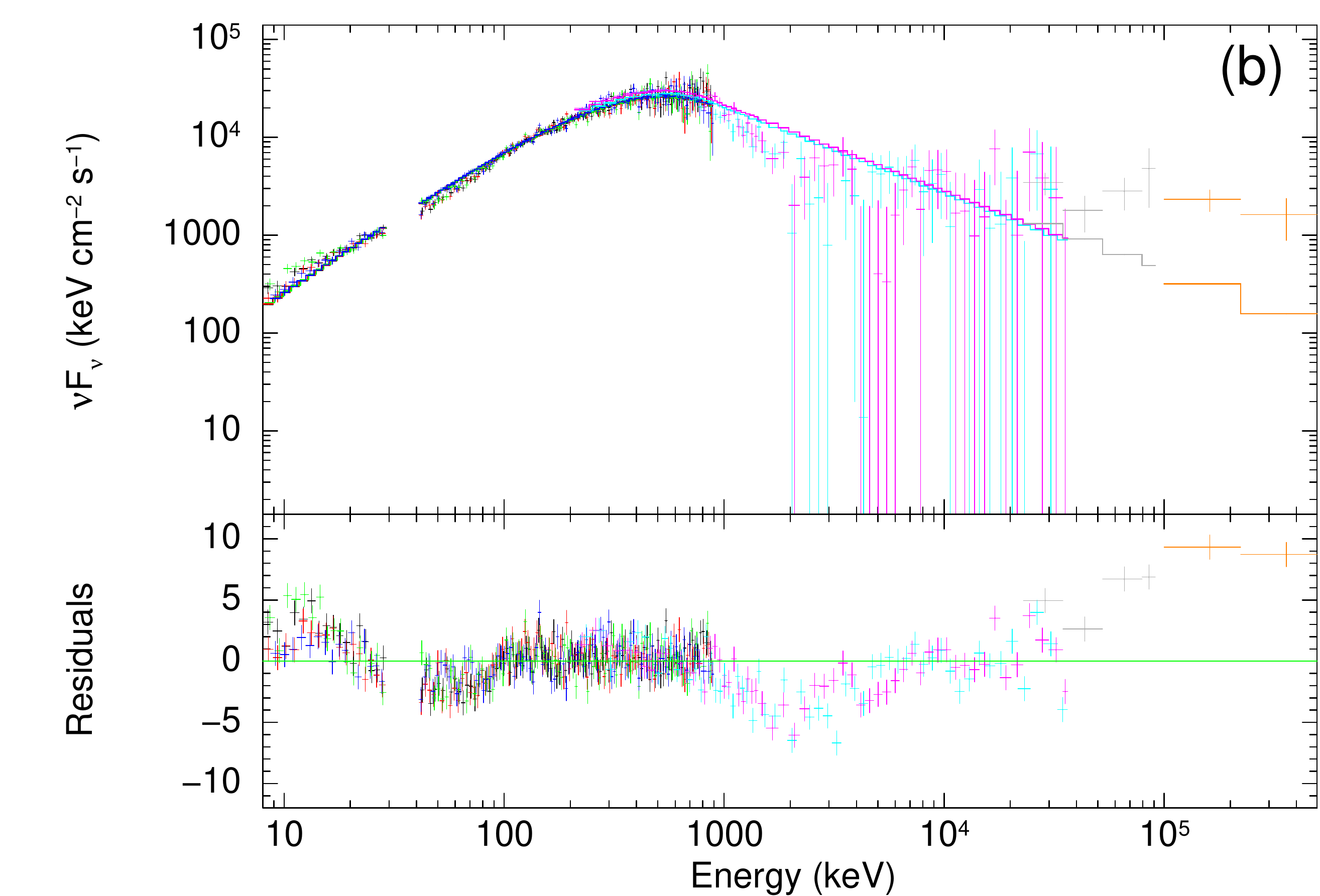}
\includegraphics[scale=0.325, angle=270]{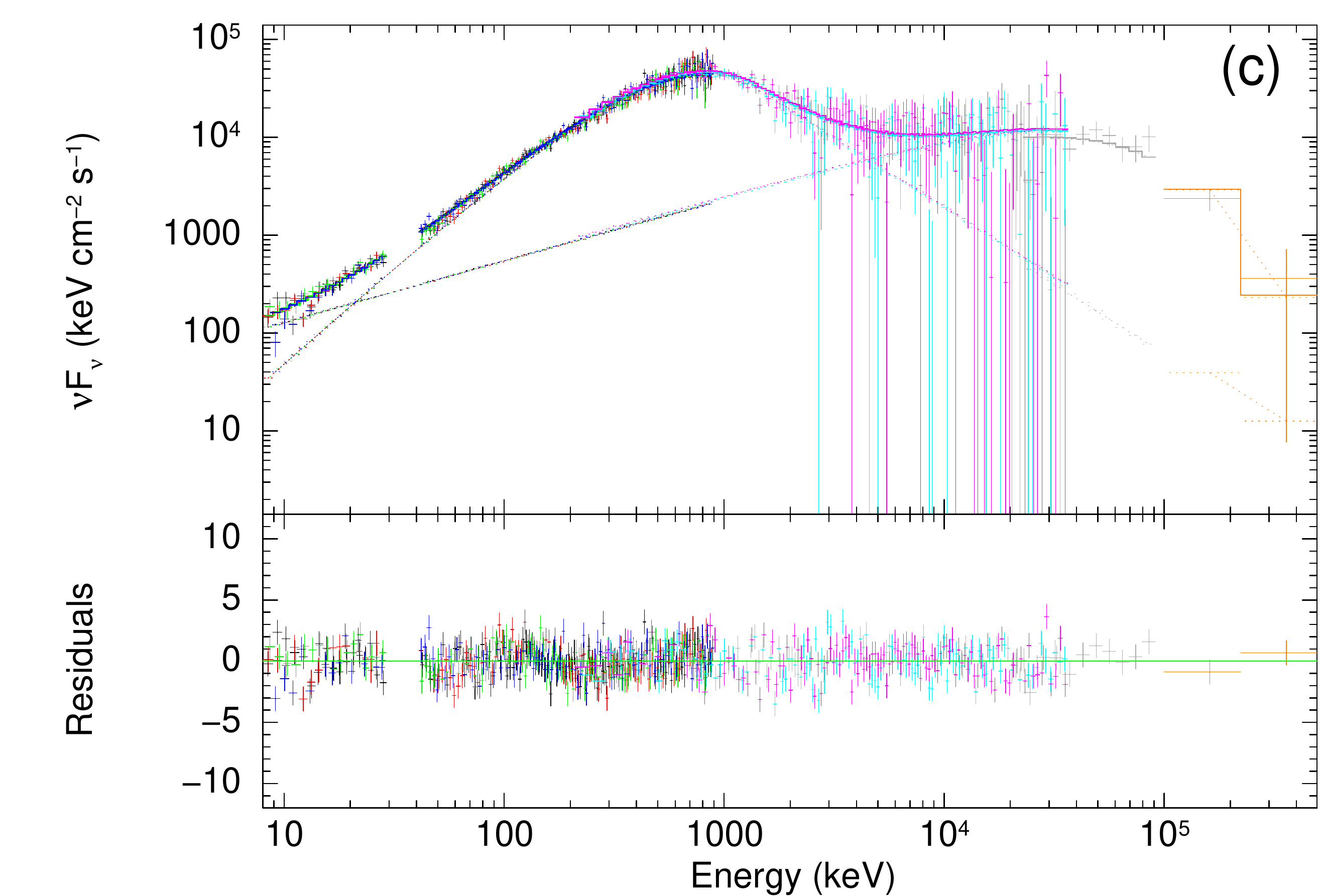}
\includegraphics[scale=0.325, angle=270]{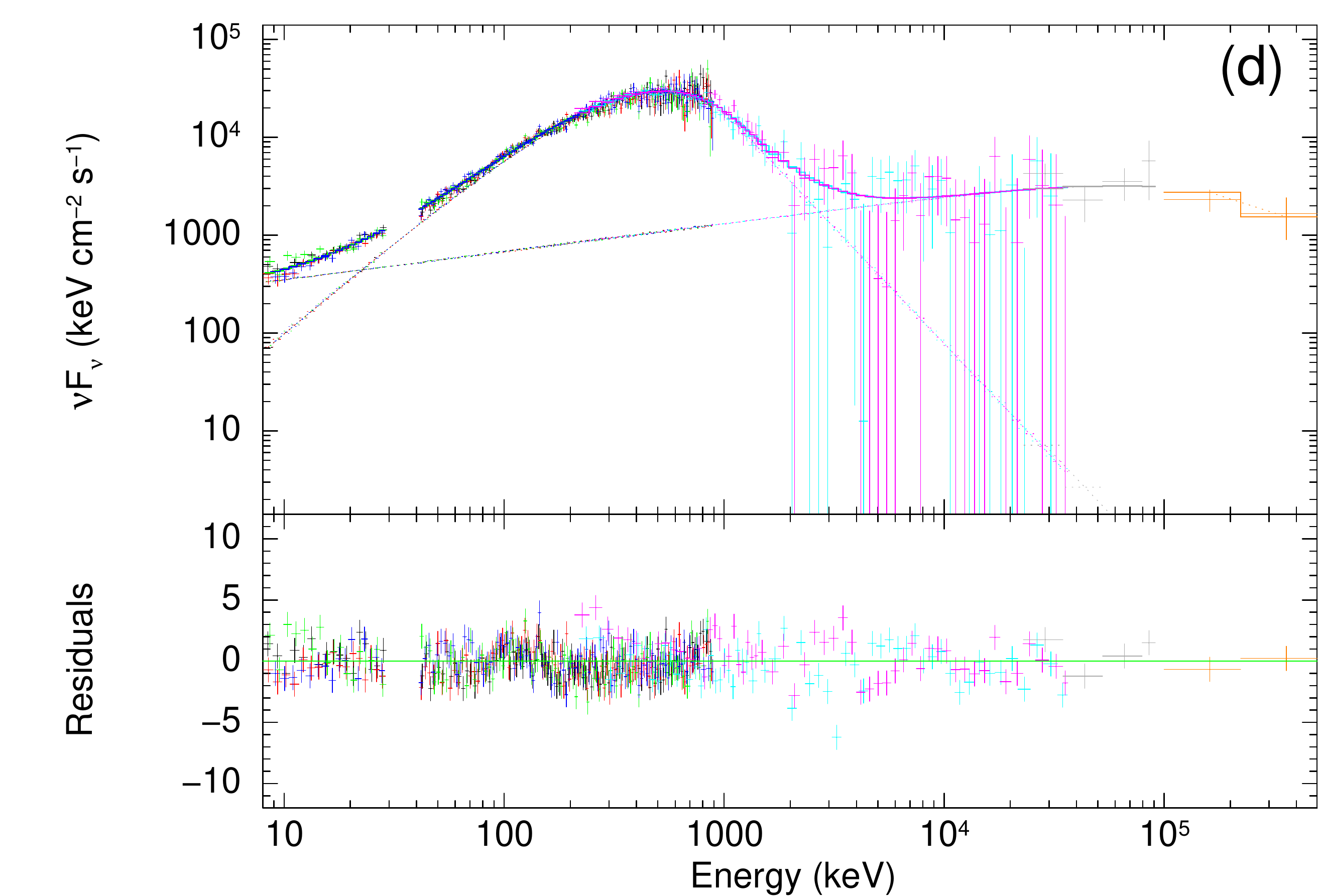}
\caption{GRB 190114C: \emph{(a)} the unfolded Band model for the $3~-~4$ $s$ interval, \emph{(b)} the unfolded Band model fitted for the 4 - 5 $s$ interval, 
\emph{(c)} the unfolded Band + cutoff power law for the 3 - 4 $s$ interval.
\emph{(d)} the unfolded Band + cutoff power law for the 4 - 5 $s$ interval.The lower panel in each plot shows the residuals. A trend can be seen in the residuals when fit with only the Band function.}
\label{fig:band45}
\end{figure*}

Looking at the light-curve morphology, it is intuitive that this cutoff, if present, will show considerable evolution as well as contamination from the afterglow. We resolve the spectrum using archived LLE data in 1 $s$ bins. Interestingly, a fit to the Band function has a systematic trend in its residuals beyond $100$ $MeV$, and this contrast is most prominent in the $3$ - $4$ $s$ bin as shown in Fig \ref{fig:band45}. This could be regarded as the signature of a cutoff in the energy spectrum around this energy range. So we added a power-law component with an exponential cutoff. The added component returned a well-constrained cutoff $\sim 50$ $MeV$ at $3$ - $4$ $s$ since the GBM trigger. The improvement in statistics strongly favors the addition of a cutoff power law. Alternatively, we modeled with BC, B + BB and BB + BC. BB + BC is strongly favoured among these; however, in comparison with B + CPL, the later is very strongly favored ($\Delta$AIC = 45, $\Delta$BIC = 44.4). So, it is evident that the high-energy data cannot be modeled by simply extrapolating the low-energy model and a cutoff is definitely required in the spectrum. 

\begin{figure}[!htb]
\centering
\includegraphics[scale=0.32, angle=270]{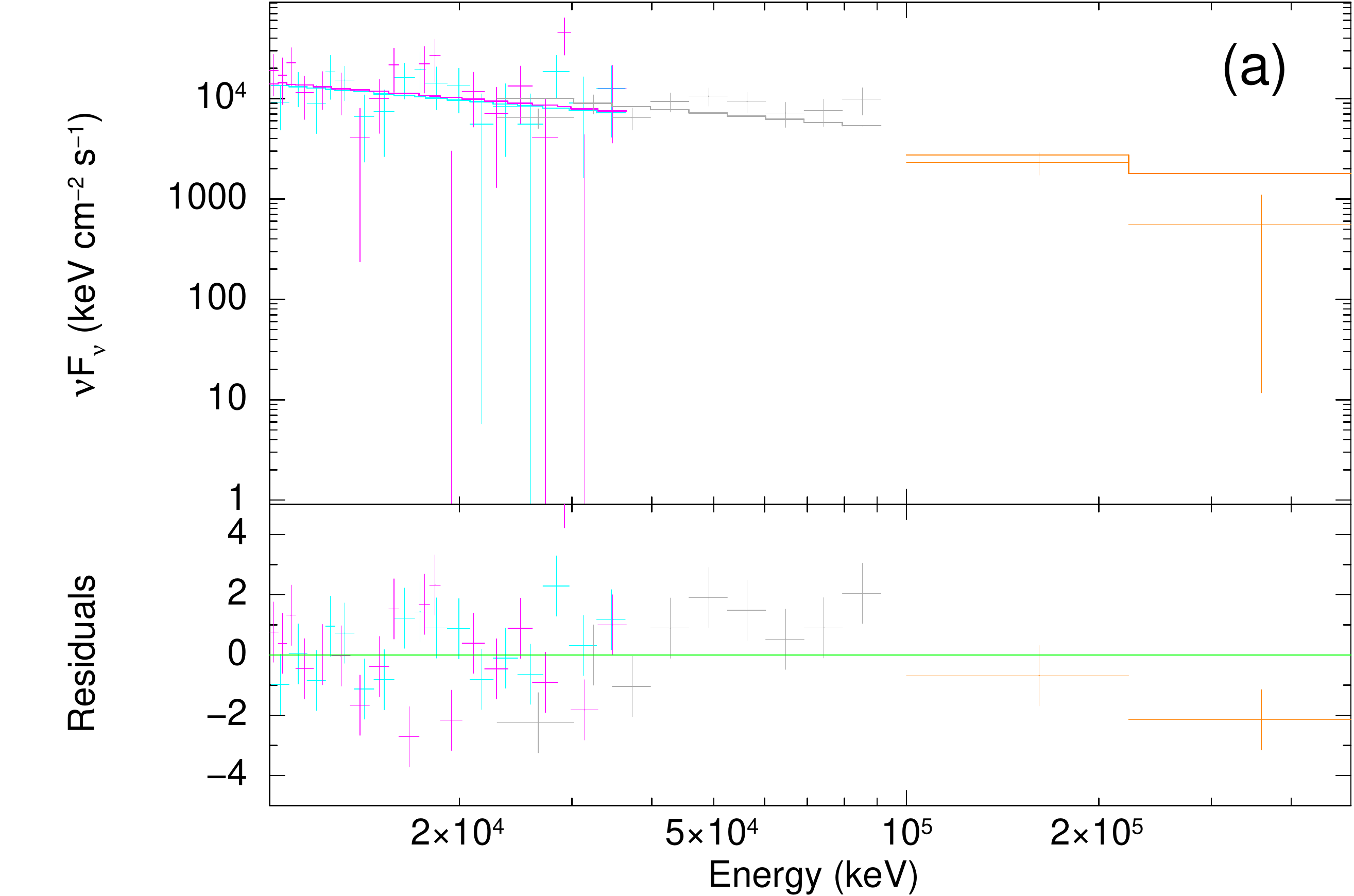} 
\includegraphics[scale=0.32, angle=270]{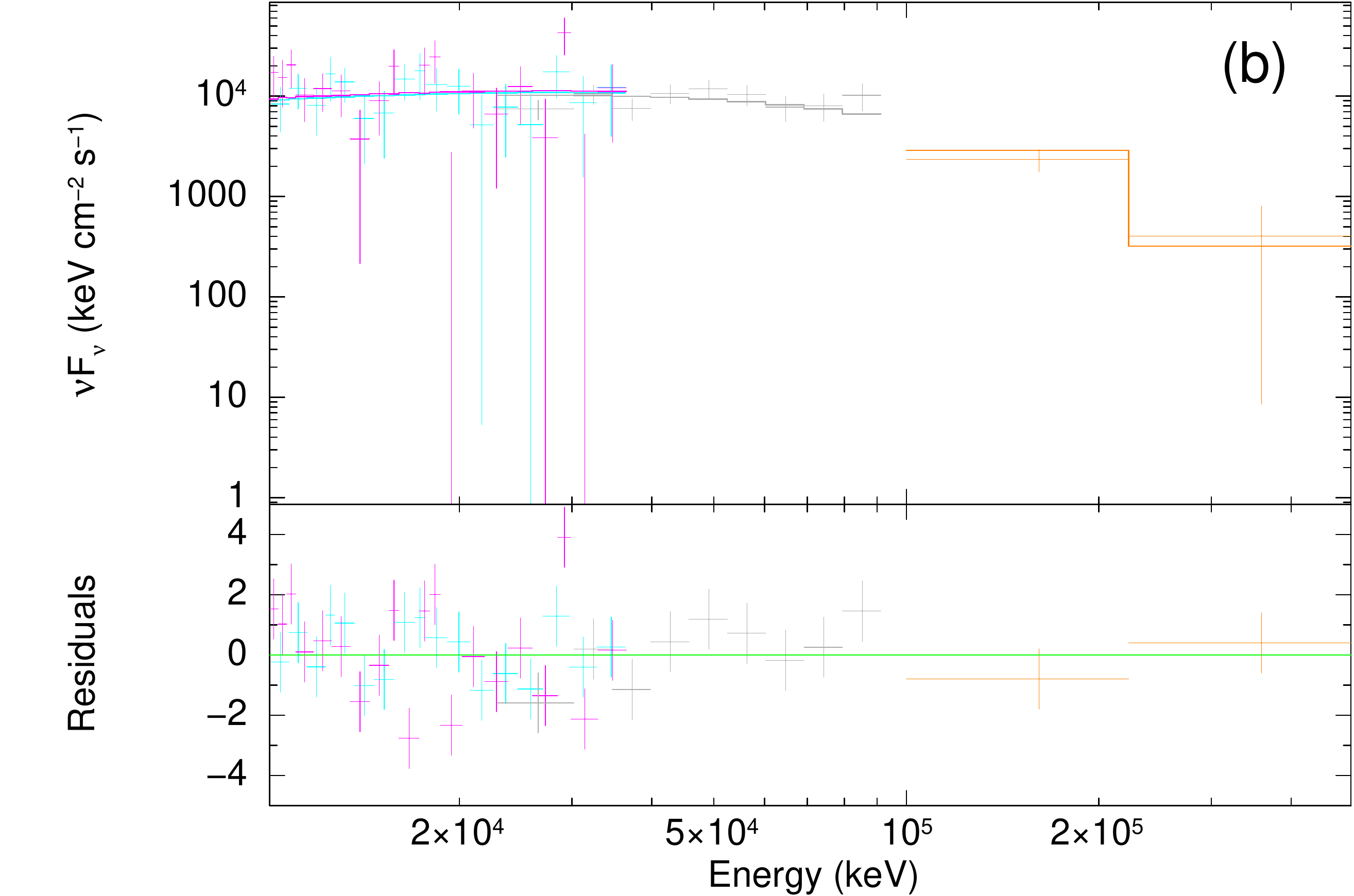}
$~~~~~~~$\includegraphics[scale=0.32, angle=270]{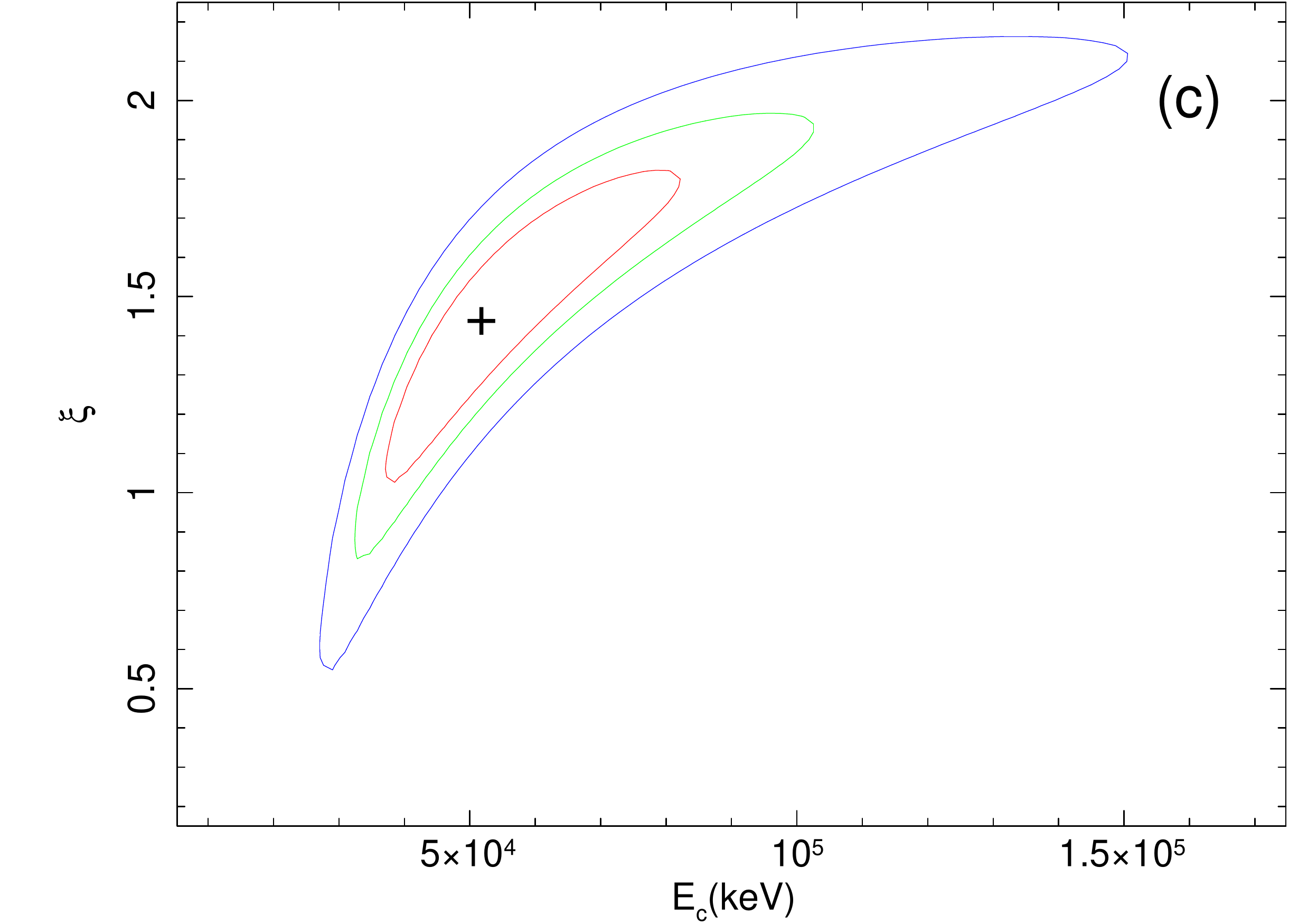}
\caption{(a) The unfolded power-law fit for energy bands above $10$ $MeV$ for the 3 - 4 $s$ interval, (b) same as the top panel but cutoff power-law fit, and (c) confidence intervals ($68\%$, $90\%$, and $99\%$) for cutoff energy and power-law index.}
\label{fig:po_cpl}
\end{figure}

The cutoff during 4-5 $s$ could not be well constrained for the upper bound. This is not surprising because of the rising contribution from the afterglow, which is significant after $4.8$ $s$ \citep{Ravasio:2019arXiv}.
The low-energy spectrum becomes harder within 3 - 5 $s$. The Band model with an additional cutoff power law fits better in these bins as reflected by the large decrease in BIC and AIC. The spectra are shown in Fig \ref{fig:band45}. Taking energies above $10$ $MeV$ we confirm that \sw{pgstat} with a cutoff power law has $20\%$ and $18\%$ contribution, respectively.
For further confirmation of the cutoff, we just take the data above $10$ $MeV$ and model it by both power law and a multiplicative cutoff. The fit to power law resulted in a slope of $2.53_{-0.13}^{+0.14}$ along with \sw{pgstat} and degree of freedom (dof)=  $107.2 (72)$, and that to cutoff power law shows a cutoff at $60_{-22}^{+70}$ $MeV$ with slope $1.5\pm0.5$ and \sw{pgstat} (dof) =  $92 (71)$. Thus, the feature could be recovered with $\Delta$BIC = $10$ which shows that an energy cutoff is very strongly preferred over a simple power-law decay. We show the fits to both models in Fig \ref{fig:po_cpl}, and also derive confidence contours for cutoff energy ($E_c$) and index ($\xi$) of the $\sim$ E$^{-\xi}$ exp(-E/$E_c$) function. The cutoff can be constrained well (Fig~\ref{fig:po_cpl}) and is also close to what is determined from the entire data fit in this interval. 

\subsubsection{Further Time-resolved analysis}


We chose a signal-to-noise ratio of 50 (from n4). 
Time bins created in this manner are reported in Table \ref{tab:specfitting}.
We chose models Band function (B), a blackbody added to band function (B + BB), a blackbody added to CPL (BB + CPL), a broken power-law model with two sharp breaks \sw{bkn2pow} and a broken power law with sharp breaks and a cutoff (\sw{bknpow}C). 
The formulas for all the models used are reported in the (Appendix \ref{sec:models}).
We presented our results in Table \ref{tab:specfitting}. During 0.7 - 1.7 $s$, \sw{bknpow}C describes the spectrum at par with BB + CPL. So, the blackbody fitted in \cite{Wang:2019arXiv} can be modeled by a low-energy break. However, in the later phase the spectrum could be modeled by BB + Band. The spectra in the later phase are also modeled with a smoothly broken power law by \cite{Ravasio:2019arXiv}, however, they also found the spectral index became harder during these times (2.45 - 5.69 $s$). Therefore, we can say that the spectrum is initially nonthermal with a low-energy cutoff (sub-$MeV$) which later becomes quasi-thermal.

 \begin{deluxetable*}{clcccccccccccc}
\tabletypesize{\scriptsize}
\tablecaption{\thisgrb: parameters of Band model and Band plus an additional cutoff power-law. \label{table_spec}}
\tablehead{
\colhead{Time} & \colhead{Model} & \colhead{$\alpha$} & \colhead{$\beta$} & \colhead{$E_{p}$}& \colhead{$K_C$}& \colhead{$\xi$\tablenotemark{a}} & \colhead{$E_{c}$} & \colhead{Norm} & \colhead{pgstat} & \colhead{AIC} & \colhead{BIC} \\
\colhead{($s$)} & \colhead{} & \colhead{} & \colhead{} & \colhead{($keV$)}& \colhead{}& \colhead{} & \colhead{($MeV$)} & \colhead{} & \colhead{(dof)} & \colhead{} & \colhead{}
}
\startdata
0-1 & B & $-0.66_{-0.02}^{+0.02}$ & $-3.08_{-0.22}^{+0.16}$ & $443_{-15}^{+16}$ & $0.53_{-0.01}^{+0.01}$ & \nodata & \nodata & \nodata  & $773.14(690)$ & $793.1$ & $838.6$\\
\hline
1-2 & B & $-0.64_{-0.01}^{+0.01}$ & $-3.94_{-0.64}^{+0.32}$ & $582_{-13}^{+13}$ & $0.76_{-0.01}^{+0.01}$ & \nodata & \nodata & \nodata & $796.17(690)$ & $816.1$ & $861.7$\\
\hline 
2-3 & B & $-0.60_{-0.01}^{+0.01}$ & $-3.46_{-0.17}^{+0.14}$ & $743_{-17}^{+10}$ & $0.52_{-0.01}^{+0.01}$ & \nodata & \nodata & \nodata & $908.11(692)$ & $928.1$ & $973.6$ \\
 & B$+$CPL & $-0.49_{-0.03}^{+0.04}$ & $-4.37_{-4.37}^{+0.49}$ & $720_{-18}^{+9}$  & $0.49_{-0.01}^{+0.01}$ & $1.66_{-0.1}^{+0.07}$ & $33_{-13}^{+47}$ & $46_{-16}^{+16}$ & $856.45(689)$ & $\underline{882.4}$ & $\underline{941.6}$ \\
\hline 
3-4 & B & $-0.26_{-0.01}^{+0.02}$ & $-2.64_{-0.03}^{+0.03}$ & $854_{-16}^{+16}$ & $0.57_{-0.01}^{+0.01}$ & \nodata & \nodata & \nodata & $1183.8(692)$ & $1203.8$ & $1249.3$ \\
 & B$+$CPL & $-0.01_{-0.04}^{+0.04}$ & $-3.42_{-0.22}^{+0.18}$ & $811_{-15}^{+16}$ & $0.48_{-0.01}^{+0.01}$ & $1.37_{-0.03}^{+0.03}$ & $50_{-8}^{+10}$ & $29_{-5}^{+5}$ & $856.02(689)$ & $\underline{882.0}$ & $\underline{941.2}$\\
\hline 
4-5 & B & $-0.46_{-0.01}^{+0.01}$ & $-2.88_{-0.05}^{+0.05}$ & $540_{-10}^{+10}$ & $0.91_{-0.01}^{+0.01}$ & \nodata & \nodata & \nodata & $1490.67(692)$ & $1510.7$ & $1556.2$ \\
 & B$+$CPL & $-0.12_{-0.03}^{+0.03}$ & $-4.03_{-0.64}^{+0.32}$ & $510_{-9}^{+9}$ & $0.83_{-0.02}^{+0.02}$ & $1.76_{-0.05}^{+0.03}$ & $474_{-177}^{+428}$ & $198_{-27}^{+23}$ & $850.88(689)$ & $\underline{876.9}$ & $\underline{936.2}$ \\
\hline 
5-6 & B & $-0.73_{-0.02}^{+0.02}$ & $-2.75_{-0.06}^{+0.06}$ & $450_{-13}^{+13}$ & $0.64_{-0.01}^{+0.01}$ & \nodata & \nodata & \nodata & $1066.99(692)$ & $1087.0$ &  $1132.5$\\
 & B$+$CPL & $-0.36_{-0.05}^{+0.05}$ & $-10.00_{-\infty}^{10.00}$ & $420_{-10}^{+11}$ & $0.60_{-0.01}^{+0.02}$ & $1.80_{-0.04}^{+0.04}$ & $183_{-71}^{+212}$ & $225_{-38}^{+41}$ & $715.85(689)$ & $\underline{741.8}$ & $\underline{801.1}$ \\
\hline 
6-7 & B & $-1.32_{-0.21}^{+0.29}$ & $-1.94_{-0.03}^{+0.03}$ & $54_{-14}^{+26}$ & $0.18_{-0.07}^{+0.2}$ & \nodata & \nodata & \nodata & $713.30(692)$ & $733.3$ & $778.8$ \\
\enddata
\end{deluxetable*}

\section{Results} 
\label{sec:conclusions}
We can calculate the $Lorentz$ factor ($\Gamma$) if the cutoff thus obtained is attributed to the $\gamma-\gamma$ absorption \citep{Lithwick:2001}, denoted here by $\Gamma_i$ (Table \ref{tab:Lorentz_factors}).
For the photons that self-annihilate, the photon energy should be comparable to the cutoff energy $E_{c}$ thus allowing us to estimate the bulk $\Gamma$ \citep{Lithwick:2001}, denoted here by $\Gamma_{ii}$.
When the Comptonization of the photons by the $e^{-}$-$e^{+}$ pairs produced by photon annihilation is also considered, the lower limit on the $\Gamma$ \citep{Lithwick:2001} is given by $\Gamma_{iii}$.
Another limit on $\Gamma$ may be obtained by considering the scattering of photons by electrons associated with baryons \citep{Lithwick:2001}, 
$\Gamma_{iv}  > {\hat \tau ^{1/\left( { - \beta  + 3} \right)}}{\left[\hat \tau ^{(-\beta-2)/(-\beta  + 3)} \times (E_{p}/m_{p}c^2)\right]^{1/5}}$,
where $E_p$ is the peak energy of the Band function and $\hat \tau$ is defined in Table \ref{tab:Lorentz_factors}. 
However, since $E_{p}/m_p c^{2}$ is for our observed $E_p$ values is $<< 1$, the above limit can be safely ignored.
Among these $\Gamma$s obtained by $\Gamma_i$ to $\Gamma_{iv}$, the highest is the relevant one. 
A larger value of $\Gamma$ is expected if the effect of 
Compton down-scattering on the cutoff energy in the comoving frame by the pair cascades is also considered \citep{Gill:2018MNRAS}.

\begin{table*}
\begin{center}
\caption{Summary of $\Gamma$s \citep{Lithwick:2001}}
\label{tab:Lorentz_factors}
\begin{tabular}{l|cccc}
\hline\hline
$Lorentz$ factor $\Gamma$& 2 - 3 s &3 - 4 s&4 - 5 s \\ \hline 

$\Gamma_{i}  = {\hat \tau ^{1/\left( { - 2\beta  + 2} \right)}}{\left( {{E_{c}}/{m_{e}}{c^2}} \right)^{\left( { - \beta  - 1} \right)/\left( { - 2\beta  + 2} \right)}}{\left( {1 + z} \right)^{\left( { - \beta  - 1} \right)/\left( { - \beta  + 1} \right)}}$, \\
where, 

$\hat \tau  = \left( {2.1 \times {{10}^{11}}} \right)\left[ {\frac{{{{\left( {d_{L}/7{{~Gpc}}} \right)}^2}{{\left( {0.511} \right)}^{\beta  + 1}}{f_{{1}}}}}{{\left( {\delta T/0.1{{~s}}} \right)\left( { - \beta  - 1} \right)}}} \right]$,
 & 115 & 211  & 318  \\ \hline

$\Gamma_{ii} \approx \frac{E_{cut}}{m_e c^2} (1+z)$
 &92 & 139 & \underline{1317}\\ \hline
 
$\Gamma_{iii}  > {\hat \tau ^{1/\left( { - \beta  + 3} \right)}}{\left( {1 + z} \right)^{\left( { - \beta  - 1} \right)/\left( { - \beta  + 3} \right)}}{\left( {180/11} \right)^{1/\left( {6 - 2\beta } \right)}} $
&\underline{154} & \underline{307} & 210\\ \hline

$E_{\rm syn, \rm max} = \frac{9\Gamma}{4\alpha_{F}(1+z)}{m_e c^2} $ $~~~~~~~~~$ in $GeV$&$6.4$ &$15.1$&$17.7$\\
\hline
\end{tabular}
\end{center}
Here, $d_{L}$ is the luminosity distance of the burst, $\beta$ is the Band high-energy index, $z$ is the redshift, $\delta T$ is the variability timescale, $f_{\rm1}$ is photon spectrum ($photon~s^{-1}~cm^{-2}~{MeV^{-1}}$) at the energy of 1 $MeV$ , and $E_{\rm{c}}$ is the cutoff energy. We extract $\delta T$ from the data by creating Bayesian blocks from the light curves by the relation $\delta T = 2t_{bb}$, $t_{bb}$ is the minimum Bayesian block and $\alpha_F$ is the fine structure constant.
\end{table*}

The various limits for the temporal bins where we could model a cutoff in the spectrum as an additional component (Table \ref{table_spec}) are calculated and the results are presented in Table \ref{tab:Lorentz_factors}. We took the time bins for our calculations that have reasonable constraints on parameters. The most relevant constraints are during the 3 - 4 $s$ time bin, where the cutoff energy $E_c/\sigma(E_c) \sim 5$, $\sigma(E_c)$ is the error in $E_c$. 
If the the emission was produced through synchrotron, then we can also estimate the limiting energy from synchrotron emission ($e.g.$ \citealt{Guilbert:1983ads, Jager:1992ApJ, Piran:2010ApJ,Atwood:2013ApJ}), which can produce a cutoff energy feature  using $E_{\rm syn, \rm max}$ defined in Table \ref{tab:Lorentz_factors}.
\citet{Fraija:2019ApJ}, through multiband spectral and temporal analysis of the afterglow using forward and reverse shocks, along with the best-fit value of the circumburst density, estimated the value of the initial bulk $\Gamma$ ($\Gamma_{0, ext}$) to be $\sim$ 600. 
The $\Gamma$ calculated from the sub-$GeV$ cutoff is $\sim 300$, which is lower than the $\Gamma$ obtained from the multi-wavelength analysis of the afterglows.

\begin{figure}[!htb]
\centering
\includegraphics[scale=0.6, angle=0]{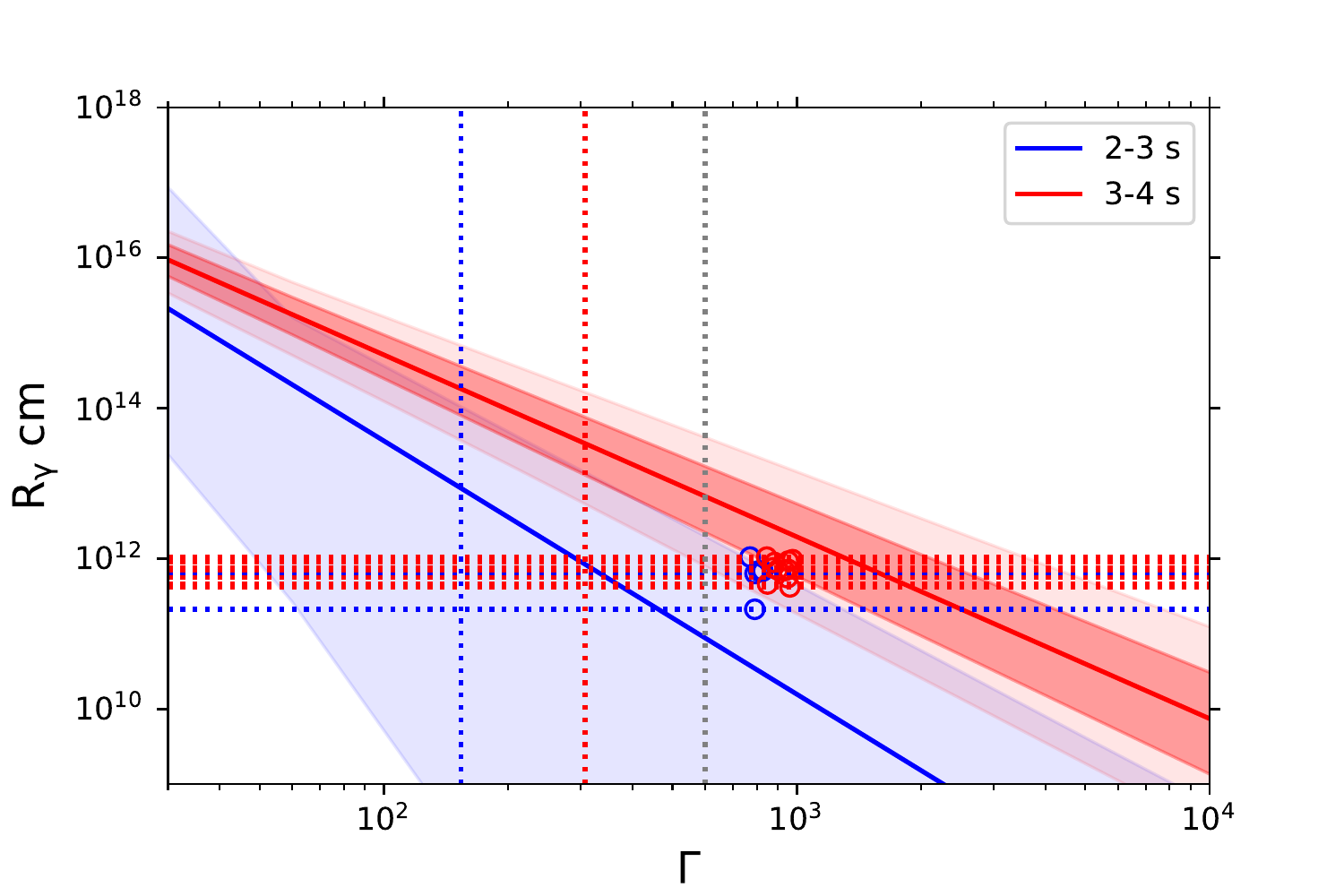}
\caption{$R_\gamma~ -~ \Gamma$ diagram for \thisgrb are shown for the labeled epochs in blue and red lines, chronologically. The allowed region lies beyond these lines. The points from the photospheric model are also shown in open circles and in the the same color same color convention. Horizontal dashed lines are used here to mark the $r_{ph}$ values. The shaded bands represent the uncertainty in the $R_\gamma ~- ~\Gamma$ region obtained by sampling the $R_\gamma$ from the observed spectral parameters taken in a range within their uncertainties. For the epoch of 3 - 4 $s$, the sampling is also performed in an interval that is twice the errorbars and shown in a lightened red shade.}
\label{fig:R_Gamma}
\end{figure}

In the internal shock model, the shock radius is inferred from the minimum variability timescale ($\rm R_{\rm IS} \sim \Gamma^2 c \delta T$). However in some models, variability timescale may not be reflection of the central engine activity \citep{Lyutikov:2003, Narayan:2009MNRAS}. For a given cutoff energy $\rm E_c$, there exists a threshold energy $E_{td}$, above which the photons annihilate with the photons at energy $\rm E_c$  and produce pairs. The threshold energy is obtained by the condition where $\rm (E_{td}/{1~{\rm
MeV}})(E_{c}/{1~{\rm MeV}}) \gtrsim 0.25 [\Gamma/(1+z)]^2$. The optical depth $\tau_{\gamma\gamma}$
is estimated depending on the relative location between $E_{td}$ and the
Band-function break energy. If energy $\rm E_{td}$ lies in between $\rm E_p(\alpha-\beta)/(2+\alpha)$ and $E_{c}$, the general expression for the optical depth is given as,

\begin{displaymath}
  \tau_{\gamma\gamma}(E)=\frac{ C(\beta) \sigma_T d_z^2 f_0}{-1-\beta}
\left(\frac{ E}{ m_e^2c^4}\right)^{-1-\beta} \frac{1}{ R_\gamma^2}
\left(\frac{\Gamma}{ 1+z}\right)^{ 2+2\beta}~,
\label{eq:tau}
\end{displaymath}

where, the parameters are as defined in \cite{Zhang:2009ApJ}. The above equation includes bulk \emph{Lorentz} factor and gamma ray emission ($R_\gamma$) radius as two independent parameters \cite{Gupta:2008MNRAS}. Therefore, we constrain the $R_\gamma - \Gamma$ diagram for the epochs 2 - 3  $s$ and 3 - 4 $s$. The best constraints are obtained for the later because of the better constrains on spectral parameters. These are shown by blue and red lines and the shaded region color coded as in Figure \ref{fig:R_Gamma}. The vertical dashed lines are the respective $\Gamma$s obtained earlier for each epoch, and the black dashed line is the $\Gamma_{0, ext}$ obtained from the blast-wave deceleration \citep{Fraija:2019ApJ}. The $R_\gamma - \Gamma$ beyond the shaded regions is allowed from the observed spectrum with sub-GeV cutoffs.

\begin{figure}[!htb]
\centering
\includegraphics[scale=0.4, angle=0]{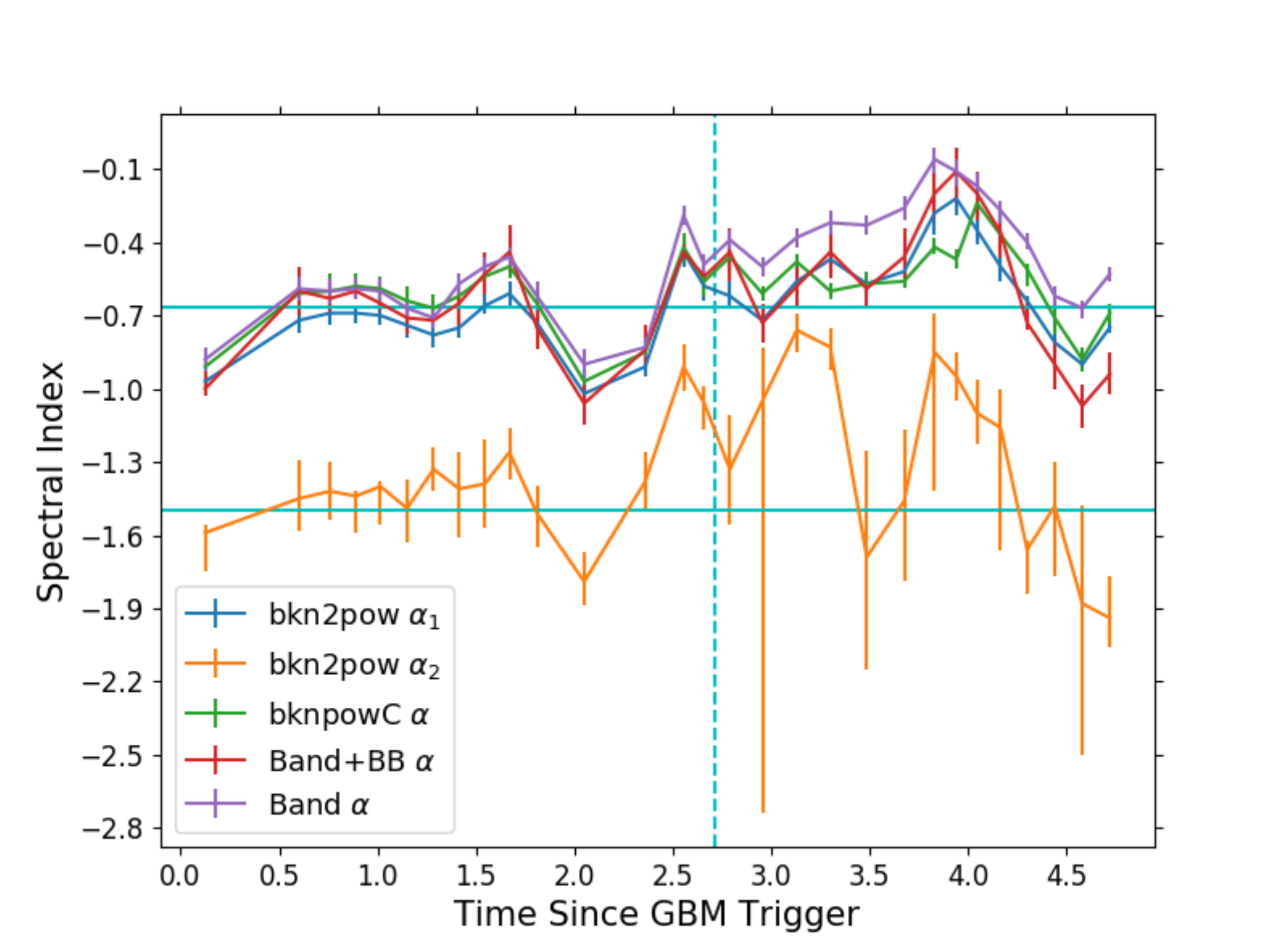}
\caption{Spectral index evolution for different models.}
\label{fig:spectral_evolution}
\end{figure}

The time-resolved spectrum also shows a transition from nonthermal to quasi-thermal. 
A blackbody added to the Band function is the preferred model just after the beginning point of the second pulse, $i.e.$, $\sim$2.5 $s$ after the GBM trigger. 
The spectral slopes are plotted in Figure \ref{fig:spectral_evolution}, where the vertical dashed line marks the beginning of the second pulse (t = 2.7 $s$). At the photosphere, the $\Gamma$ of the GRB outflow is found from the calculation as given in the \citep{Asaf:2007ApJ}.

\begin{displaymath}
\label{eq:lorentz_Asaf}
  \Gamma_{ph}  = {\left[ {\left( {1.06} \right){{\left( {1 + z} \right)}^2}{d_{\rm{L}}}\frac{{Y{F_{\gamma {\rm{,ob}}}}{\sigma _{\rm{T}}}}}{{2{m_{\rm{p}}}{c^3}{\cal R}}}} \right]^{1/4}},
\end{displaymath}

where observed energy flux is $F_{\gamma {\rm{,ob}}}$, $Y$ is the ratio of the total burst energy to the energy emitted in $\gamma$-rays and ${\cal R}$ is defined as
${\cal R} \equiv {\left( {\frac{{{F_{{\rm{bb,ob}}}}}}{{\sigma T_{{\rm{ob}}}^4}}} \right)^{1/2}}$,
such that $F_{\rm bb,ob}$ and $T_{\rm ob}$ are the observed flux and temperature of the blackbody component, respectively. The relevant radii, the initial fireball radius ($r_0$), the saturation radius ($r_s$) and the photospheric radius ($r_{ph}$) are defined in \citealt{Asaf:2007ApJ}.
The efficiency for \thisgrb is taken from \cite{Fraija:2019ApJ} for calculating the Lorentz factor ($Y$ = 6.65 corresponding to $15\%$ efficiency). The $\Gamma$s thus determined are reported in Table \ref{tab:specfitting} and are also shown by open circles in Figure \ref{fig:R_Gamma} with colors representing the epochs.

The hard spectrum during the second pulse may suggest dissipation below the photosphere. The outgoing photons produced in this dissipation undergo process Comptonization and escape at the photosphere, which is a fuzzy ($i.e.$, extended) zone \citep{Beloborodov:2017rev}. The specialty of such a model spectrum is its strikingly similar shape as the observed one during this time, and it can account for both the hardness as well as the high-energy cutoff seen in the spectrum. In this scenario, the radiation is emitted at the photospheric radius $r_{ph}$. The $\Gamma$ thus obtained is similar to the $\Gamma_{ph}$, and within possible uncertainties in the $\Gamma_{0, ext}$. A consistent picture is thus obtained where the prompt emission spectrum showing a sub-$GeV$ cutoff is arising from a subphotospheric zone.

\section{Discussion} 
\label{sec:discussion}
In the literature, several bright GRBs have shown a peculiar cutoff in the sub-$GeV$ energies \citep{Tang:2015ApJ, Wang:2017, Vianello:2017}. These GRBs are observed in a broadband. The analysis of the prompt emission for many GRBs could be marred by limited band observations and also by contamination from the bright afterglow component existing simultaneously with prompt emission. 
We have shown this here in the case of \thisgrb (also seen in \citealt{Ravasio:2019arXiv}), and in the case of \oldgrb, we highlighted the transition region by showing the evolution of prompt emission and afterglows. In the case of \oldgrb, the effect is feeble while \thisgrb is severely affected.
In our analysis of the joint data from \fermi detectors, we have recovered such a break in the time-resolved analysis that could be smeared in the time-integrated emission and at a certain point by the dominating external component. 

\thisgrb also shows a spectacular evolution of the spectral shape. 
Some bursts with well-separated phases have shown a transition from a fireball dominant outflow to a Poynting flux dominated outflow \citep{Zhang:2018Nat, Li:2019ApJS}. Some bursts show a hybrid of thermal and nonthermal components e.g., GRB 100724B, GRB 110721A, and GRB 120323A \citep{Axelsson:2012ApJ}. In the case of GRB 190114C, initially the emission is nonthermal and can be a synchrotron emission from internal shock with a weak photospheric component. The initial spectrum becomes hard after $\sim$ 2.5 $s$ where a fresh injection seems to be occurring. Such a hard spectrum can be possible in the case of dissipation occurring  at a region below the photosphere, which is optically thick ($\tau>1$). By this time the first pulse contribution is also subdominant. The shape of the spectrum can be explained in such a case arising from the Comptonization of the outgoing photons. The observed sub-$GeV$ cutoff in the spectrum helps constrain the site of the prompt emission observed in \thisgrb. The complete picture from this analysis would be as follows: (a) the initial phase is produced away from the photosphere, (b) the second hard phase is produced in a subphotospheric dissipation, and (c) the afterglow is produced in the external shock; the sub-$TeV$ radiation is also consistent with this picture arising from inverse-Compton scattering \citep{Fraija:2019ApJ, MAGIC2019Natur_IC} in the external shocks.

\section*{Acknowledgments}
We are thankful for and acknowledge helpful discussions with Jagdish C. Joshi. We thank Nestor Mirabel of the Fermi Help Desk for his extensive support during this project. V.C., P.S.P., X.B.H. and P.H.T. are supported by the National Natural Science Foundation of China (NSFC) grants
11633007, 11661161010, and U1731136. P.S.P. acknowledges sysu-postdoctoral fellowship.
This research has made use of data obtained through the HEASARC Online Service, provided by the 
NASA-GSFC, in support of NASA High Energy Astrophysics Programs.

\begin{table*}
\caption{Time-resolved Spectral fitting for different models  }

\label{tab:specfitting}
\begin{tiny}
\begin{center}
\begin{tabular}{c|ccccccllll}
\hline
B&&&  && & $dof = 560$\\
\hline
Sr. no. &(t$_1$,t$_2$) & $\alpha$ & $\beta$ & $E_p$ ($keV$) & $K_{B}$ &AIC & BIC & BIC & AIC model  \\
        &&&&&&&                                                                & model (B, BB + B) & preferred model\\
\hline
1&($-0.26$ , $0.51 $)  &$-0.88_{-0.05}^{+0.05}$ & $-2.84_{-1.34}^{+0.34}$ & $466_{-53}^{+60}$ & $0.16_{-0.01}^{+0.01}$ & 693.0 & 727.8      & B & \sw{bknpow}C, BB + CPL, \sw{bkn2pow}, BB + B, B\\
2&($0.51 $ , $0.69 $)  &$-0.59_{-0.05}^{+0.05}$ & $-3.0_{-0.5}^{+0.3}$ & $417_{-30}^{+31}$ & $0.71_{-0.04}^{+0.05}$ & 556.0 & 590.7      & B & B, \sw{bkn2pow}, BB + B, \sw{bknpow}C \\
3&($0.69 $ , $0.83 $)  &$-0.60_{-0.05}^{+0.05}$ & $-2.9_{-0.4}^{+0.2}$ & $439_{-32}^{+34}$ & $0.86_{-0.05}^{+0.06}$  & 559.1 & 593.8     & B, BB + B & \sw{bkn2pow}, BB + B, B  \\
4&($0.83 $ , $0.95 $)  &$-0.59_{-0.04}^{+0.04}$ & $-3.8_{-\infty}^{+0.6}$ & $465_{-27}^{+28}$ & $0.95_{-0.05}^{+0.05}$  & 510.9 & 545.6   & B & \sw{bknpow}C, BB + CPL, BB + B, B, \sw{bkn2pow} \\
5&($0.95 $ , $1.08 $)  &$-0.60_{-0.05}^{+0.05}$ & $-3.36_{-1.15}^{+0.42}$ & $423_{-27}^{+29}$ & $0.97_{-0.05}^{+0.06}$   & 564.0 & 598.7     & B & \sw{bknpow}C, BB + B, BB + CPL, B\\
6&($1.08 $ , $1.22 $)  &$-0.67_{-0.04}^{+0.04}$ & $-3.5_{-2.8}^{+0.4}$ & $504_{-35}^{+36}$ & $0.84_{-0.04}^{+0.05}$ & 507.7 & 542.5      & B, BB + B & \sw{bknpow}C, BB + CPL, BB + B\\
7&($1.22 $ , $1.35 $)  &$-0.71_{-0.03}^{+0.04}$ & $-9.4_{-\infty}^{+5.1}$ & $656_{-38}^{+40}$ & $0.74_{-0.03}^{+0.03}$  & 514.8 & 549.6   & B & \sw{bknpow}C, B, BB + CPL, \sw{bkn2pow}, BB + B\\
8&($1.35 $ , $1.48 $)  &$-0.57_{-0.04}^{+0.04}$ & $-4_{-3.4}^{+0.5}$ & $645_{-32}^{+34}$ & $0.84_{-0.03}^{+0.04}$ & 569.1 & 603.8      & B , BB + B & BB + B, \sw{bkn2pow}, BB + CPL, \sw{bknpow}C, B\\
9&($1.48 $ , $1.61 $)  &$-0.50_{-0.04}^{+0.04}$ & $-4.1_{-\infty}^{+0.6}$ & $593_{-29}^{+31}$ & $0.89_{-0.04}^{+0.04}$    & 473.3 & 508.1  &B & B, \sw{bknpow}C, BB + B, BB + CPL\\
10&($1.61 $ , $1.73 $) &$-0.46_{-0.04}^{+0.04}$ & $-5.6_{-\infty}^{+4.1}$ & $727_{-30}^{+33}$ & $0.88_{-0.04}^{+0.04}$  & 602.9 & 637.6   & B & B, \sw{bknpow}C, BB + CPL, BB + B\\
11&($1.73 $ , $1.89 $) &$-0.62_{-0.06}^{+0.06}$ & $-2.70_{-0.31}^{+0.18}$ & $346_{-32}^{+36}$ & $0.84_{-0.07}^{+0.08}$  & 577.3 & 612.0     & BB + B, B & \sw{bknpow}C, BB + B, BB + CPL, \sw{bkn2pow}\\
12&($1.89 $ , $2.21 $) &$-0.90_{-0.06}^{+0.07}$ & $-2.58_{-0.53}^{+0.23}$ & $317_{-43}^{+51}$ & $0.37_{-0.03}^{+0.05}$  & 580.8 & 615.5     & B, BB + B & \sw{bknpow}C, BB + CPL, \sw{bkn2pow}, BB + B\\
13&($2.21 $ , $2.51 $) &$-0.83_{-0.04}^{+0.04}$ & $-9.4_{-\infty}^{+0.9}$ & $676_{-74}^{+56}$ & $0.34_{-0.02}^{+0.02}$  & 545.2 & 579.9   & B & \sw{bkn2pow}, B, \sw{bknpow}C, BB + CPL, BB + B\\
14&($2.51 $ , $2.60 $) &$-0.29_{-0.04}^{+0.04}$ & $-4.3_{-\infty}^{+0.6}$ & $742_{-30}^{+32}$ & $1.16_{-0.05}^{+0.05}$  & 552.6 & 587.4   & B & \sw{bkn2pow}, B, \sw{bknpow}C, BB + CPL, BB + B\\
15&($2.60 $ , $2.72 $) &$-0.49_{-0.04}^{+0.04}$ & $-3.6_{-0.6}^{+0.3}$ & $763_{-36}^{+40}$ & $0.81_{-0.03}^{+0.03}$ & 628.5 & 663.2      & B & \sw{bkn2pow}, B\\
16&($2.72 $ , $2.86 $) &$-0.39_{-0.04}^{+0.04}$ & $-4.9_{-\infty}^{+1.1}$ & $678_{-27}^{+31}$ & $0.72_{-0.03}^{+0.03}$  & 568.6 & 603.4   & B, BB + B &\sw{bknpow}C, BB + B, BB + CPL\\
17&($2.86 $ , $3.05 $) &$-0.50_{-0.04}^{+0.04}$ & $-3.4_{-0.4}^{+0.3}$ & $823_{-41}^{+44}$ & $0.51_{-0.02}^{+0.02}$  & 628.5 & 663.2     & BB + B & BB + B, \sw{bkn2pow}\\
18&($3.05 $ , $3.21 $) &$-0.38_{-0.04}^{+0.04}$ & $-2.66_{-0.11}^{+0.10}$ & $963_{-49}^{+52}$ & $0.54_{-0.02}^{+0.02}$  & 696.1 & 730.9     & BB + B & BB + B, \sw{bkn2pow}\\
19&($3.21 $ , $3.39 $) &$-0.32_{-0.05}^{+0.05}$ & $-2.47_{-0.09}^{+0.08}$ & $835_{-45}^{+48}$ & $0.51_{-0.02}^{+0.02}$ & 613.6 & 648.3      & BB + B & BB + B\\
20&($3.39 $ , $3.58 $) &$-0.33_{-0.04}^{+0.04}$ & $-3.24_{-0.27}^{+0.20}$ & $939_{-41}^{+40}$ & $0.51_{-0.02}^{+0.02}$   & 646.8 & 681.5    & BB + B & BB + B\\
21&($3.58 $ , $3.77 $) &$-0.26_{-0.05}^{+0.05}$ & $-2.93_{-0.17}^{+0.14}$ & $877_{-40}^{+41}$ & $0.46_{-0.02}^{+0.02}$ & 731.6 & 766.3      & BB + B & BB + B\\
22&($3.77 $ , $3.89 $) &$-0.06_{-0.05}^{+0.06}$ & $-2.69_{-0.11}^{+0.09}$ & $815_{-37}^{+39}$ & $0.72_{-0.03}^{+0.03}$ & 660.3 & 695.0      & BB + B & BB + B\\
23&($3.89 $ , $3.99 $) &$-0.11_{-0.05}^{+0.06}$ & $-2.66_{-0.11}^{+0.10}$ & $690_{-34}^{+35}$ & $0.98_{-0.04}^{+0.05}$  & 661.8 & 696.6     & B , BB + B & \sw{bkn2pow}\\
24&($3.99 $ , $4.10 $) &$-0.17_{-0.05}^{+0.05}$ & $-3.5_{-0.4}^{+0.3}$ & $668_{-27}^{+29}$ & $0.97_{-0.04}^{+0.04}$ & 643.4 & 678.2      & B & \sw{bkn2pow}, BB + B\\
25&($4.10 $ , $4.23 $) &$-0.27_{-0.04}^{+0.05}$ & $-3.37_{-0.42}^{+0.27}$ & $607_{-28}^{+27}$ & $0.88_{-0.04}^{+0.04}$ & 641.7 & 676.4      & BB + B & BB + B\\
26&($4.23 $ , $4.37 $) &$-0.40_{-0.04}^{+0.03}$ & $-3.29_{-0.41}^{+0.26}$ & $412_{-21}^{+7}$ & $1.03_{-0.05}^{+0.02}$   & 581.5 & 616.3     & BB + B & \sw{bkn2pow}, BB + B\\
27&($4.37 $ , $4.51 $) &$-0.62_{-0.04}^{+0.05}$ & $-3.42_{-\infty}^{+0.50}$ & $424_{-26}^{+27}$ & $0.88_{-0.05}^{+0.05}$   & 627.8 & 662.6  & BB + B & \sw{bkn2pow}, BB + B, BB + CPL\\
28&($4.51 $ , $4.65 $) &$-0.67_{-0.03}^{+0.04}$ & $-4.3_{-\infty}^{+0.7}$ & $520_{-29}^{+11}$ & $0.83_{-0.03}^{+0.02}$  & 653.0 & 687.7   & BB + B & BB + CPL, BB + B\\
29&($4.65 $ , $4.79 $) &$-0.53_{-0.03}^{+0.03}$ & $-4.34_{-\infty}^{+0.76}$ & $615_{-12}^{+12}$ & $0.77_{-0.01}^{+0.01}$  & 671.0 & 705.7   & BB + B & BB + CPL, BB + B\\
\hline

\end{tabular}
\\ $b$: $photons~ keV^{-1} ~ cm^{-2} ~s^{-1}$ at 1 keV
\end{center}
\end{tiny}
\end{table*}

\addtocounter{table}{-1}
\begin{table*}
\caption{Time-resolved Spectral fitting (continued)}
\begin{tiny}
\begin{center}
\begin{tabular}{c|ccccccccccll}
\hline
B + BB&&&   &&&&&  $dof=558$ &\\
\hline 
Sr. no. &(t$_1$,t$_2$) &$\alpha$& $\beta$ &  $E_p$  & $K_{B}$      & $kT_{BB}$  & $K_{BB}$                               & AIC & BIC &$\Gamma_{ph}$& $r_{ph}$ & $r_{s}$ \\ 
Sr. no. & in $s$ & &  &  ($keV$)  &     &  ($keV$)  &                               & & & (Y = 6.65)& ($10^{11}$ $cm$) & ($10^{11}$ $cm$) \\ 
\hline
1&($-0.26$ , $0.51 $) 	&$-1.0_{-0.07}^{+0.03}$			&$-3.6_{-\infty}^{+0.9}$		&$609_{-91}^{+93}$		&$0.12_{-0.02}^{+0.03}$ &$43_{-6}^{+6}$			&$9.5_{-3.4}^{+5.7}$	&692.0 		&735.4	&\nodata &\nodata&\nodata	\\
2&($0.51 $ , $0.69 $) 	&$-0.6_{-0.1}^{+0.1}$			&$-3.14_{-0.74}^{+0.33}$		&$449_{-78}^{+63}$		&$0.63_{-0.10}^{+0.11}$ &$40_{-40}^{+\infty}$	&$19_{-19}^{+33}$		&558.0 		&601.4 	&\nodata&\nodata	&\nodata\\
3&($0.69 $ , $0.83 $) 	&$-0.63_{-0.07}^{+0.07}$		&$-3.1_{-0.51}^{+0.3}$			&$488_{-45}^{+52}$		&$0.74_{-0.08}^{+0.09}$ &$33_{-10}^{+11}$		&$28_{-18}^{+25}$		&555.3 		&598.7	&\nodata&\nodata&\nodata	\\
4&($0.83 $ , $0.95 $) 	&$-0.60_{-0.04}^{+0.03}$		&$-4.0_{-\infty}^{+0.7}$		&$490_{-23}^{+11}$		&$0.86_{-0.07}^{+0.02}$ &$32_{-6}^{+8}$			&$23_{-6.6}^{+19}$		&509.8 		&553.3	&\nodata&\nodata&\nodata	\\
5&($0.95 $ , $1.08 $) 	&$-0.65_{-0.08}^{+0.03}$		&$-3.65_{-2.86}^{+0.50}$		&$463_{-38}^{+11}$		&$0.83_{-0.11}^{+0.06}$ &$39_{-10}^{+7}$		&$36_{-20}^{+37}$		&562.6 		&606.0	&\nodata&\nodata&\nodata	\\
6&($1.08 $ , $1.22 $) 	&$-0.71_{-0.07}^{+0.07}$		&$-3.9_{-\infty}^{+0.8}$		&$559_{-52}^{+61}$		&$0.72_{-0.08}^{+0.08}$ &$36_{-12}^{+11}$		&$34_{-22}^{+29}$		&502.5 		&545.9	&\nodata&\nodata&\nodata	\\
7&($1.22 $ , $1.35 $) 	&$-0.72_{-0.04}^{+0.03}$		&$-10_{-\infty}^{+10}$			&$671_{-17}^{+25}$		&$0.71_{-0.05}^{+0.16}$ &$35_{-35}^{+\infty}$	&$12_{-12}^{+19}$		&517.0		&560.6	&\nodata&\nodata&\nodata\\
8&($1.35 $ , $1.48 $) 	&$-0.65_{-0.08}^{+0.09}$		&$-3.7_{-1.9}^{+0.9}$			&$620_{-233}^{+84}$		&$0.76_{-0.08}^{+0.11}$ &$158_{-38}^{+49}$		&$216_{-122}^{+337}$	&564.2 		&607.6	&\nodata&\nodata&\nodata\\
9&($1.48 $ , $1.61 $) 	&$-0.53_{-0.09}^{+0.08}$		&$-3.8_{-4.5}^{+0.9}$			&$574_{-214}^{+76}$		&$0.86_{-0.10}^{+0.13}$ &$160_{-92}^{+129}$		&$104_{-104}^{+398}$	&475.8 		&519.2	&\nodata&\nodata&\nodata	\\
10&($1.61 $ , $1.73 $)	&$-0.44_{-0.11}^{+0.08}$		&$-3.5_{-\infty}^{+0.7}$		&$495_{-130}^{+246}$	&$0.94_{-0.14}^{+0.11}$ &$239_{-239}^{+760}$	&$608_{-608}^{+259}$	&604.9 		&648.3	&\nodata&\nodata &\nodata\\
11&($1.73 $ , $1.89 $)	&$-0.75_{-0.09}^{+0.09}$		&$-3.15_{-\infty}^{+0.41}$		&$451_{-58}^{+63}$		&$0.60_{-0.08}^{+0.10}$ &$33_{-7.1}^{+6.5}$		&$41_{-20}^{+23}$		&567.7 		&611.1	&\nodata&\nodata &\nodata\\
12&($1.89 $ , $2.21 $)	&$-1.06_{-0.08}^{+0.09}$		&$-3.2_{-\infty}^{+0.6}$		&$455_{-77}^{+91}$		&$0.26_{-0.04}^{+0.05}$ &$32_{-7}^{+6}$			&$18_{-9}^{+9}$			&574.2 		&617.6	&\nodata&\nodata &\nodata	\\
13&($2.21 $ , $2.51 $)	&$-0.84_{-0.10}^{+0.07}$		&$-3.73_{-\infty}^{+0.98}$		&$654_{-252}^{+161}$	&$0.33_{-0.05}^{+0.04}$ &$156_{-124}^{+\infty}$	&$23_{-23}^{+128}$		&547.9 		&591.4	&\nodata& \nodata &\nodata\\
14&($2.51 $ , $2.60 $)	&$-0.44_{-0.03}^{+0.03}$		&$-10.0_{-\infty}^{+10.0}$		&$797_{-17}^{+17}$		&$0.97_{-0.02}^{+0.02}$ &$153_{-10}^{+11}$		&$455_{-40}^{+41}$		&550.4 		&593.8	&846 	&12.0	&  0.98 	\\
15&($2.60 $ , $2.72 $)	&$-0.54_{-0.03}^{+0.03}$		&$-10_{-\infty}^{+10}$			&$758_{-17}^{+17}$		&$0.77_{-0.01}^{+0.01}$ &$201_{-27}^{+32}$		&$169_{-33}^{+33}$		&641.2 		&684.7	&962 	&11.6	&  0.91 	\\
16&($2.72 $ , $2.86 $)	&$-0.44_{-0.10}^{+0.12}$		&$-3.1_{-3.6}^{+0.6}$			&$466_{-173}^{+229}$	&$0.71_{-0.10}^{+0.15}$ &$195_{-45}^{+23}$		&$536_{-374}^{+252}$	&562.3 		&605.7	&828 	&11.8	& 1.00 	\\
17&($2.86 $ , $3.05 $)	&$-0.73_{-0.08}^{+0.08}$		&$-2.67_{-0.40}^{+0.38}$		&$735_{-277}^{+191}$	&$0.40_{-0.04}^{+0.05}$ &$184_{-19}^{+19}$		&$448_{-106}^{+183}$	&575.2 		&618.6	&808 	&11.8	&  0.99	\\
18&($3.05 $ , $3.21 $)	&$-0.58_{-0.08}^{+0.08}$		&$-2.30_{-0.15}^{+0.18}$		&$929_{-214}^{+8849}$	&$0.44_{-0.04}^{+0.04}$ &$208_{-18}^{+23}$		&$573_{-146}^{+184}$	&657.0 		&700.4	&954 	&11.8	&  0.98	\\
19&($3.21 $ , $3.39 $)	&$-0.44_{-0.08}^{+0.11}$		&$-2.06_{-0.16}^{+0.17}$		&$621_{-199}^{+177}$	&$0.46_{-0.04}^{+0.07}$ &$205_{-19}^{+16}$		&$524_{-165}^{+178}$	&585.0 		&628.4	&935 	&11.8	&  0.97	\\
20&($3.39 $ , $3.58 $)	&$-0.59_{-0.03}^{+0.07}$		&$-2.78_{-0.20}^{+0.22}$		&$1000_{-143}^{+8937}$	&$0.39_{-0.01}^{+0.03}$ &$192_{-9}^{+19}$		&$546_{-29}^{+126}$		&607.3 		&650.7	&875 	&11.8	&  0.99	\\
21&($3.58 $ , $3.77 $)	&$-0.46_{-0.12}^{+0.09}$		&$-2.22_{-0.33}^{+0.16}$		&$616_{-138}^{+358}$	&$0.39_{-0.06}^{+0.04}$ &$212_{-30}^{+13}$		&$682_{-213}^{+131}$	&690.2 		&733.6	&899 	&11.8	& 1.00	\\
22&($3.77 $ , $3.89 $)	&$-0.2_{-0.1}^{+0.1}$			&$-2.14_{-0.17}^{+0.14}$		&$584_{-111}^{+146}$	&$0.64_{-0.07}^{+0.07}$ &$211_{-14}^{+14}$		&$1054_{-272}^{+216}$	&632.3 		&675.7	&977 	&11.9	& 1.00	\\
23&($3.89 $ , $3.99 $)	&$-0.11_{-0.10}^{+0.13}$		&$-2.25_{-0.24}^{+0.17}$		&$479_{-116}^{+146}$	&$1.00_{-0.11}^{+0.17}$ &$200_{-14}^{+22}$		&$822_{-428}^{+274}$	&655.7 		&699.1	&954 	&11.9	&  0.99	\\
24&($3.99 $ , $4.10 $)	&$-0.20_{-0.09}^{+0.11}$		&$-2.7_{-0.6}^{+0.3}$			&$475_{-124}^{+201}$	&$0.96_{-0.11}^{+0.14}$ &$194_{-43}^{+25}$		&$774_{-565}^{+299}$	&641.2 		&684.7	&883 	&11.9	& 1.01	\\
25&($4.10 $ , $4.23 $)	&$-0.36_{-0.10}^{+0.10}$		&$-2.43_{-0.29}^{+0.19}$		&$381_{-77}^{+113}$		&$0.83_{-0.10}^{+0.11}$ &$161_{-9}^{+12}$		&$675_{-198}^{+134}$	&625.6 		&669.0	&771 	&12.0	& 1.02	\\
26&($4.23 $ , $4.37 $)	&$-0.73_{-0.10}^{+0.03}$		&$-3.05_{-0.52}^{+0.27}$		&$524_{-58}^{+17}$		&$0.60_{-0.09}^{+0.07}$ &$74_{-7}^{+4}$			&$212_{-60}^{+13}$		&563.1 		&606.6	&507 	&12.2	& 1.01	\\
27&($4.37 $ , $4.51 $)	&$-0.9_{-0.1}^{+0.1}$			&$-2.83_{-0.65}^{+0.31}$		&$511_{-83}^{+87}$		&$0.56_{-0.09}^{+0.11}$ &$75_{-8.2}^{+13.1}$	&$170_{-51}^{+48}$		&607.7 		&651.1	&507 	&12.2 	& 1.00	\\
28&($4.51 $ , $4.65 $)	&$-1.07_{-0.09}^{+0.09}$		&$-3.6_{-\infty}^{+0.8}$		&$621_{-106}^{+134}$	&$0.47_{-0.06}^{+0.07}$ &$105_{-7.4}^{+8.8}$	&$356_{-56}^{+58}$		&567.0 		&610.4	&582 	&12.1 	& 1.03	\\
29&($4.65$ , $4.79$)	&$-0.94_{-0.09}^{+0.08}$		&$-3.73_{-\infty}^{+0.73}$		&$837_{-105}^{+139}$ 	&$0.44_{-0.05}^{+0.06}$ &$114_{-7}^{+8}$		&$416_{-61}^{+76}$ 		&603.9 		&647.3	&628 	&12.1	& 1.02	\\

\hline
\end{tabular}
\\ $b$: $photons~ keV^{-1} ~ cm^{-2} ~s^{-1}$ at 1 keV
\end{center}
\end{tiny}
\end{table*}

\addtocounter{table}{-1}
\begin{table*}
\caption{Time-resolved Spectral fitting (continued)}
\begin{tiny}
\begin{center}
\begin{tabular}{c|ccccccccccD}
\hline
BB+CPL&&&   &&&&&  $dof=559$ &\\
\hline 
Sr. no. &(t$_1$,t$_2$) & $\xi$ & $E_{c}$ &$K_{C}$ &    $kT_{BB}$   & $K_{BB}$ &  AIC & BIC\\ \hline 
1&($-0.26$ , $0.51 $) &$1.00_{-0.08}^{+0.08}$&$619_{-132}^{+199}$&$12.49_{-2.80}^{+3.62}$&$43_{-10}^{+10}$&$9_{-6}^{+6}$&          690.5 & 729.6 \\
2&($0.51 $ , $0.69 $) &$0.70_{-0.09}^{+0.07}$&$381_{-29}^{+28}$&$13.81_{-3.47}^{+3.17}$&$45_{-4.}^{+7}$&$36_{-6}^{+24}$&          562.0 & 601.1 \\
3&($0.69 $ , $0.83 $) &$0.66_{-0.06}^{+0.07}$&$393_{-43}^{+58}$&$15.09_{-3.37}^{+4.31}$&$35_{-9}^{+10}$&$37_{-19}^{+26}$&          563.1 & 602.2 \\
4&($0.83 $ , $0.95 $) &$0.60_{-0.07}^{+0.04}$&$355_{-19}^{+22}$&$14.12_{-3.36}^{+2.28}$&$32_{-6}^{+8}$&$24_{-7}^{+20}$&            509.8 & 548.9 \\
5&($0.95 $ , $1.08 $) &$0.67_{-0.07}^{+0.08}$&$359_{-36}^{+53}$&$17.73_{-3.74}^{+5.25}$&$40_{-9}^{+8}$&$42_{-20}^{+38}$&          563.0 & 602.1 \\
6&($1.08 $ , $1.22 $) &$0.73_{-0.06}^{+0.06}$&$446_{-49}^{+63}$&$19.51_{-4.25}^{+5.02}$&$36_{-11}^{+10}$&$38_{-21}^{+27}$&       501.1 & 540.2 \\
7&($1.22 $ , $1.35 $) &$0.72_{-0.04}^{+0.04}$&$523_{-42}^{+52}$&$19.45_{-3.28}^{+3.65}$&$34_{-18}^{+\infty}$&$11_{-12}^{+19}$&       515.1 & 554.2 \\
8&($1.35 $ , $1.48 $) &$0.66_{-0.08}^{+0.08}$&$479_{-99}^{+84}$&$15.80_{-3.76}^{+5.11}$&$150_{-34}^{+55}$&$199_{-112}^{+145}$&    565.6 & 604.7 \\
9&($1.48 $ , $1.61 $) &$0.55_{-0.08}^{+0.09}$&$419_{-86}^{+68}$&$10.49_{-2.70}^{+3.77}$&$125_{-125}^{+\infty}$&$77_{-77}^{+160}$&    476.5 & 515.6 \\
10&($1.61 $ , $1.73 $)&$0.48_{-0.09}^{+0.07}$&$442_{-162}^{+78}$&$7.96_{-2.03}^{+2.47}$&$242_{-129}^{+223}$&$198_{-198}^{+545}$&  603.4 & 642.5 \\
11&($1.73 $ , $1.89 $)&$0.77_{-0.07}^{+0.07}$&$390_{-50}^{+65}$&$20.09_{-4.56}^{+5.78}$&$34_{-6}^{+6}$&$45_{-18}^{+21}$&            567.9 & 607.0 \\
12&($1.89 $ , $2.21 $)&$1.07_{-0.08}^{+0.08}$&$505_{-98}^{+145}$&$35.6_{-7.79}^{+9.79}$&$32_{-7}^{+6}$&$18_{-9}^{+9}$&           573.0 & 612.1 \\
13&($2.21 $ , $2.51 $)&$0.86_{-0.07}^{+0.08}$&$631_{-148}^{+134}$&$16.81_{-3.78}^{+5.00}$&$65_{-33}^{+\infty}$&$13_{-13}^{+53}$&        546.1 & 585.2 \\
14&($2.51 $ , $2.60 $)&$0.44_{-0.09}^{+0.09}$&$512_{-59}^{+75}$&$7.49_{-2.15}^{+2.95}$&$152_{-20}^{+29}$&$455_{-240}^{+225}$&     548.4 & 587.5 \\
15&($2.60 $ , $2.72 $)&$0.54_{-0.05}^{+0.07}$&$520_{-86}^{+65}$&$9.21_{-1.91}^{+2.65}$&$200_{-61}^{+85}$&$169_{-87}^{+211}$&      639.3 & 678.4 \\
16&($2.72 $ , $2.86 $)&$0.48_{-0.08}^{+0.09}$&$431_{-110}^{+67}$&$6.02_{-1.57}^{+2.20}$&$170_{-34}^{+52}$&$239_{-136}^{+236}$&     563.1 & 602.2 \\
17&($2.86 $ , $3.05 $)&$0.80_{-0.08}^{+0.08}$&$808_{-135}^{+179}$&$14.17_{-3.57}^{+4.71}$&$164_{-12}^{+14}$&$390_{-82}^{+80}$&    594.7 & 633.7 \\
18&($3.05 $ , $3.21 $)&$0.94_{-0.06}^{+0.07}$&$2393_{-405}^{+515}$&$21.74_{-4.34}^{+5.17}$&$172_{-8}^{+8}$&$817_{-100}^{+100}$&      741.0 & 780.1 \\
19&($3.21 $ , $3.39 $)&$0.96_{-0.07}^{+0.06}$&$2793_{-609}^{+831}$&$19.55_{-4.33}^{+5.11}$&$152_{-7}^{+7}$&$666_{-85}^{+82}$&        744.3 & 783.4 \\
20&($3.39 $ , $3.58 $)&$0.71_{-0.08}^{+0.08}$&$1038_{-164}^{+207}$&$8.69_{-2.34}^{+3.02}$&$171_{-10}^{+11}$&$577_{-118}^{+114}$&  634.7 & 673.8 \\
21&($3.58 $ , $3.77 $)&$0.76_{-0.09}^{+0.09}$&$1365_{-284}^{+384}$&$9.69_{-2.81}^{+3.71}$&$163_{-8}^{+8}$&$615_{-106}^{+104}$&      764.2 & 803.3 \\
22&($3.77 $ , $3.89 $)&$0.84_{-0.08}^{+0.07}$&$2161_{-456}^{+566}$&$13.42_{-3.55}^{+4.24}$&$158_{-7}^{+6}$&$1278_{-152}^{+142}$&      758.1 & 797.2 \\
23&($3.89 $ , $3.99 $)&$0.70_{-0.13}^{+0.10}$&$1188_{-315}^{+400}$&$11.06_{-4.06}^{+4.99}$&$123_{-9}^{+7}$&$858_{-214}^{+179}$&     808.6 & 847.7 \\
24&($3.99 $ , $4.10 $)&$0.32_{-0.12}^{+0.11}$&$444_{-115}^{+77}$&$3.45_{-1.19}^{+1.73}$&$121_{-45}^{+41}$&$244_{-215}^{+202}$&     658.9 & 698.0 \\
25&($4.10 $ , $4.23 $)&$0.46_{-0.11}^{+0.10}$&$443_{-63}^{+80}$&$5.85_{-1.85}^{+2.63}$&$121_{-17}^{+26}$&$274_{-140}^{+130}$&       649.0 & 688.0 \\
26&($4.23 $ , $4.37 $)&$0.73_{-0.10}^{+0.09}$&$451_{-77}^{+52}$&$19.15_{-5.49}^{+6.90}$&$73_{-6}^{+3}$&$221_{-61}^{+12}$&         569.5 & 608.6 \\
27&($4.37 $ , $4.51 $)&$0.86_{-0.09}^{+0.09}$&$444_{-73}^{+100}$&$31.33_{-8.45}^{+11.32}$&$74_{-9}^{+12}$&$147_{-52}^{+49}$&     612.2 & 651.3 \\
28&($4.51 $ , $4.65 $)&$1.07_{-0.08}^{+0.08}$&$687_{-138}^{+190}$&$65.61_{-15.73}^{+20.22}$&$104_{-7}^{+8}$&$355_{-56}^{+57}$&     565.2 & 604.3 \\
29&($4.65 $ , $4.79 $)&$0.95_{-0.01}^{+0.01}$&$810_{-40}^{+44}$&$34.01_{-1.23}^{+0.95}$&$113_{-3}^{+4.0}$&$416_{-20}^{+18}$&          602.7 & 641.8 \\

\hline

\end{tabular}
\end{center}
\end{tiny}
\end{table*}


\addtocounter{table}{-1}
\begin{table*}
\caption{Time-resolved Spectral fitting (continued)}
\begin{tiny}
\begin{center}
\begin{tabular}{c|ccccccccccD}
\hline
\sw{bknpow}C&&&   &&&&&  $dof=559$ &\\
\hline 
Sr. no. &(t$_1$,t$_2$) & $\alpha_1$      &   $E_1$ ($keV$)      &   $\alpha_2$          & $K$ &  $E_{c}$ & AIC & BIC\\ \hline
1&($-0.26$ , $0.51 $) & $ 0.91_{-0.04}^{+0.04}$ & $137.17_{-17.3}^{+20.83}$ & $1.35_{-0.23}^{+0.23}$ & $9.61_{-1.28}^{+1.38}$ &  $892.85_{-209.79}^{+554.67}$ &		 	687.4 & 726.5 \\
2&($0.51 $ , $0.69 $) & $ 0.61_{-0.03}^{+0.03}$ & $129.62_{-13.24}^{+10.17}$ & $0.97_{-0.21}^{+0.2}$ & $11.02_{-1.34}^{+1.42}$ &  $447.91_{-46.01}^{+101.7}$ & 			559.8 & 598.9 \\
3&($0.69 $ , $0.83 $) & $ 0.6_{-0.07}^{+0.06}$ & $110.42_{-30.16}^{+27.47}$ & $0.96_{-0.09}^{+0.19}$ & $12.69_{-2.81}^{+3.05}$ &  $482.62_{-85.63}^{+131.31}$ & 		561.6 & 600.7 \\
4&($0.83 $ , $0.95 $) & $ 0.58_{-0.05}^{+0.03}$ & $114.67_{-29.45}^{+25.36}$ & $0.79_{-0.18}^{+0.15}$ & $13.43_{-1.8}^{+1.63}$ &  $398.66_{-35.14}^{+55.02}$ & 			510.2 & 549.3 \\
5&($0.95 $ , $1.08 $) & $ 0.59_{-0.05}^{+0.05}$ & $114.43_{-23.51}^{+35.77}$ & $0.89_{-0.17}^{+0.21}$ & $14.53_{-2.54}^{+2.82}$ &  $406.72_{-52.53}^{+91.89}$ & 		561.8 & 600.9 \\
6&($1.08 $ , $1.22 $) & $ 0.64_{-0.06}^{+0.06}$ & $107.77_{-61.34}^{+41.31}$ & $0.95_{-0.15}^{+0.19}$ & $16.07_{-3.2}^{+3.9}$ &  $522.58_{-81.46}^{+127.46}$ & 			500.1 & 539.2 \\
7&($1.22 $ , $1.35 $) & $ 0.67_{-0.06}^{+0.06}$ & $324.51_{-97.93}^{+96.88}$ & $0.28_{-0.43}^{+0.39}$ & $17.51_{-3.39}^{+5.43}$ &  $391.94_{-75.27}^{+214.55}$ & 		513.4 & 552.5 \\
8&($1.35 $ , $1.48 $) & $ 0.62_{-0.05}^{+0.03}$ & $743.43_{-235.58}^{+152.87}$ & $1.47_{-0.55}^{+0.58}$ & $13.81_{-2.49}^{+3.07}$ &  $557.75_{-79.58}^{+178.53}$ & 		566.6 & 605.7 \\
9&($1.48 $ , $1.61 $) & $ 0.54_{-0.08}^{+0.06}$ & $475.3_{-\infty}^{+\infty}$ & $0.81_{-0.69}^{+0.48}$ & $9.98_{-2.11}^{+2.66}$ &  $455.75_{-126.86}^{+146.15}$ & 		476.7 & 515.8 \\
10&($1.61 $ , $1.73 $)& $ 0.5_{-0.08}^{+0.05}$ & $781.97_{-\infty}^{+\infty}$ & $0.94_{-0.51}^{+0.48}$ & $8.44_{-1.67}^{+1.84}$ &  $531.33_{-142.04}^{+87.4}$ & 		602.9 & 642.0 \\
11&($1.73 $ , $1.89 $)& $ 0.65_{-0.06}^{+0.06}$ & $101.37_{-15.44}^{+24.15}$ & $1.13_{-0.09}^{+0.19}$ & $14.66_{-2.85}^{+3.53}$ &  $497.77_{-92.12}^{+142.0}$ & 		566.5 & 605.6 \\
12&($1.89 $ , $2.21 $)& $ 0.97_{-0.05}^{+0.05}$ & $112.34_{-21.18}^{+19.34}$ & $1.47_{-0.22}^{+0.2}$ & $27.19_{-4.51}^{+5.16}$ &  $762.17_{-246.11}^{+517.69}$ & 		571.9 & 610.9 \\
13&($2.21 $ , $2.51 $)& $ 0.85_{-0.05}^{+0.05}$ & $621.38_{-\infty}^{+\infty}$ & $1.23_{-0.81}^{+0.59}$ & $16.39_{-3.22}^{+3.53}$ &  $670.21_{-140.12}^{+224.96}$ & 	545.8 & 584.9 \\
14&($2.51 $ , $2.60 $)& $ 0.42_{-0.06}^{+0.06}$ & $583.31_{-57.78}^{+66.95}$ & $1.39_{-0.45}^{+0.44}$ & $6.88_{-1.52}^{+1.86}$ &  $719.25_{-154.81}^{+263.92}$ & 		541.4 & 580.5 \\
15&($2.60 $ , $2.72 $)& $ 0.56_{-0.05}^{+0.06}$ & $716.86_{-122.16}^{+\infty}$ & $1.24_{-0.47}^{+0.48}$ & $9.74_{-1.93}^{+2.4}$ &  $669.58_{-117.41}^{+204.59}$ & 		635.7 & 674.8 \\
16&($2.72 $ , $2.86 $)& $ 0.46_{-0.06}^{+0.06}$ & $687.69_{-118.26}^{+182.38}$ & $1.3_{-0.5}^{+0.51}$ & $5.56_{-1.18}^{+1.47}$ &  $540.5_{-84.18}^{+134.19}$ & 			562.2 & 601.3 \\
17&($2.86 $ , $3.05 $)& $ 0.61_{-0.03}^{+0.03}$ & $659.96_{-68.66}^{+79.95}$ & $1.75_{-0.21}^{+0.27}$ & $7.4_{-0.97}^{+1.1}$ &  $1000.0_{-75.62}^{+\infty}$ & 			603.2 & 642.3 \\
18&($3.05 $ , $3.21 $)& $ 0.48_{-0.03}^{+0.03}$ & $591.72_{-85.88}^{+93.09}$ & $1.03_{-0.11}^{+0.13}$ & $4.47_{-0.66}^{+0.74}$ &  $1000.0_{-386.19}^{+\infty}$ & 		836.4 & 875.5 \\
19&($3.21 $ , $3.39 $)& $ 0.6_{-0.03}^{+0.03}$ & $534.91_{-34.13}^{+34.66}$ & $2.38_{-0.07}^{+0.07}$ & $5.84_{-0.91}^{+1.01}$ &  $--_{---}^{---}$ & 					618.5 & 657.6 \\
20&($3.39 $ , $3.58 $)& $ 0.57_{-0.05}^{+0.04}$ & $617.75_{-38.23}^{+39.99}$ & $2.49_{-0.45}^{+0.35}$ & $5.48_{-1.07}^{+1.18}$ &  $7184.06_{-4767.06}^{+\infty}$&       632.7 & 671.8 \\
21&($3.58 $ , $3.77 $)& $ 0.56_{-0.03}^{+0.03}$ & $588.7_{-31.56}^{+21.41}$ & $2.65_{-0.13}^{+0.06}$ & $4.4_{-0.69}^{+0.56}$ &  $---_{---}^{---}$ & 					729.7 & 768.8 \\
22&($3.77 $ , $3.89 $)& $ 0.42_{-0.04}^{+0.03}$ & $534.85_{-26.59}^{+25.82}$ & $2.51_{-0.07}^{+0.07}$ & $3.54_{-0.61}^{+0.69}$ &  $---_{---}^{---}$ & 					689.2 & 728.3 \\
23&($3.89 $ , $3.99 $)& $ 0.47_{-0.04}^{+0.04}$ & $458.14_{-26.61}^{+28.35}$ & $2.46_{-0.14}^{+0.09}$ & $5.78_{-0.97}^{+1.15}$ &  $---_{---}^{---}$ & 	                730.0 & 769.1 \\
24&($3.99 $ , $4.10 $)& $ 0.24_{-0.08}^{+0.07}$ & $557.71_{-\infty}^{+\infty}$ & $0.58_{-0.39}^{+0.48}$ & $2.7_{-0.69}^{+0.89}$ &  $437.47_{-77.19}^{+115.05}$ & 		660.6 & 699.7 \\
25&($4.10 $ , $4.23 $)& $ 0.37_{-0.08}^{+0.07}$ & $436.99_{-97.63}^{+158.91}$ & $0.98_{-0.47}^{+0.47}$ & $4.25_{-1.03}^{+1.32}$ &  $501.82_{-108.93}^{+189.98}$ & 		652.5 & 691.6 \\
26&($4.23 $ , $4.37 $)& $ 0.51_{-0.02}^{+0.07}$ & $213.93_{-29.5}^{+10.02}$ & $1.21_{-0.16}^{+0.4}$ & $8.87_{-1.33}^{+1.89}$ &  $471.94_{-97.74}^{+139.7}$ & 			582.9 & 622.0 \\
27&($4.37 $ , $4.51 $)& $ 0.71_{-0.07}^{+0.07}$ & $313.09_{-107.63}^{+145.24}$ & $1.34_{-0.47}^{+0.52}$ & $19.69_{-4.26}^{+5.27}$ &  $471.09_{-123.53}^{+237.63}$ & 	624.7 & 663.8 \\
28&($4.51 $ , $4.65 $)& $ 0.88_{-0.05}^{+0.05}$ & $417.43_{-33.02}^{+42.94}$ & $2.77_{-0.44}^{+0.43}$ & $35.19_{-6.22}^{+6.78}$ &  $3269.48_{-1994.58}^{+\infty}$ & 	596.8 & 635.9 \\
29&($4.65 $ , $4.79 $)& $ 0.69_{-0.04}^{+0.04}$ & $424.23_{-38.77}^{+14.82}$ & $2.04_{-0.37}^{+0.23}$ & $14.96_{-2.35}^{+2.83}$ &  $1327.91_{-451.3}^{+81.85}$ & 		640.5 & 679.6 \\

\hline
\sw{bkn2pow}&&& &  &&&&$dof = 558$\\
\hline 
Sr. no. &(t$_1$,t$_2$) & $\alpha_1$      &   $E_1$ ($keV$)      &   $\alpha_2$          & $E_2$ (keV) &  $\beta$  & $K^b$ & AIC & BIC\\ \hline    
1&($-0.26$ , $0.51 $) &$0.97_{-0.03}^{+0.03}$&$132_{-11}^{+16}$&$1.59_{-0.03}^{+0.16}$&$557_{-86}^{+234}$&$2.79_{-0.18}^{+0.51}$&$11.08_{-1.40}^{+2.02}$&  690.5 & 733.9 \\
2&($0.51 $ , $0.69 $) &$0.72_{-0.05}^{+0.05}$&$126_{-21}^{+17}$&$1.45_{-0.16}^{+0.13}$&$464_{-71}^{+85}$&$2.93_{-0.24}^{+0.35}$&$14.80_{-2.78}^{+3.15}$&     556.7 & 600.1 \\
3&($0.69 $ , $0.83 $) &$0.69_{-0.06}^{+0.05}$&$114_{-18}^{+16}$&$1.42_{-0.12}^{+0.12}$&$503_{-60}^{+71}$&$2.92_{-0.20}^{+0.28}$&$16.28_{-3.40}^{+3.81}$&    553.7 & 597.1 \\
4&($0.83 $ , $0.95 $) &$0.69_{-0.03}^{+0.04}$&$125_{-9}^{+18}$&$1.44_{-0.02}^{+0.15}$&$563_{-33}^{+90}$&$3.36_{-0.17}^{+0.45}$&$18.25_{-2.51}^{+2.68}$&     512.1 & 555.5 \\
5&($0.95 $ , $1.08 $) &$0.70_{-0.04}^{+0.04}$&$115_{-11}^{+18}$&$1.40_{-0.02}^{+0.16}$&$446_{-46}^{+75}$&$2.94_{-0.11}^{+0.26}$&$19.38_{-3.12}^{+3.38}$&    570.0 & 613.5 \\
6&($1.08 $ , $1.22 $) &$0.74_{-0.06}^{+0.05}$&$127_{-21}^{+24}$&$1.49_{-0.12}^{+0.14}$&$640_{-78}^{+113}$&$3.29_{-0.29}^{+0.45}$&$20.99_{-4.20}^{+4.77}$&     507.7 & 551.1 \\
7&($1.22 $ , $1.35 $) &$0.78_{-0.05}^{+0.05}$&$135_{-25}^{+24}$&$1.33_{-0.09}^{+0.09}$&$754_{-69}^{+79}$&$3.68_{-0.32}^{+0.45}$&$23.41_{-4.33}^{+4.90}$&     515.8 & 559.3 \\
8&($1.35 $ , $1.48 $) &$0.75_{-0.05}^{+0.04}$&$227_{-53}^{+71}$&$1.41_{-0.15}^{+0.20}$&$752_{-69}^{+68}$&$3.61_{-0.30}^{+0.38}$&$20.09_{-3.65}^{+3.77}$&     565.5 & 608.9 \\
9&($1.48 $ , $1.61 $) &$0.66_{-0.05}^{+0.03}$&$181_{-37}^{+46}$&$1.39_{-0.18}^{+0.18}$&$703_{-104}^{+94}$&$3.57_{-0.37}^{+0.49}$&$14.05_{-2.57}^{+2.14}$&  484.8 & 528.2 \\
10&($1.61 $ , $1.73 $)&$0.61_{-0.05}^{+0.05}$&$190_{-33}^{+42}$&$1.26_{-0.10}^{+0.11}$&$795_{-53}^{+71}$&$3.57_{-0.23}^{+0.31}$&$11.39_{-2.20}^{+2.49}$&     611.1 & 654.5 \\
11&($1.73 $ , $1.89 $)&$0.73_{-0.06}^{+0.05}$&$102_{-11}^{+17}$&$1.51_{-0.11}^{+0.14}$&$424_{-57}^{+110}$&$2.77_{-0.18}^{+0.30}$&$18.12_{-3.52}^{+4.29}$&    570.7 & 614.1 \\
12&($1.89 $ , $2.21 $)&$1.02_{-0.04}^{+0.04}$&$115_{-13}^{+14}$&$1.79_{-0.12}^{+0.10}$&$630_{-176}^{+177}$&$3.17_{-0.56}^{+1.04}$&$31.29_{-4.94}^{+5.61}$&     573.4 & 616.8 \\
13&($2.21 $ , $2.51 $)&$0.91_{-0.04}^{+0.04}$&$158_{-36}^{+38}$&$1.38_{-0.12}^{+0.12}$&$681_{-74}^{+131}$&$3.13_{-0.27}^{+0.42}$&$19.04_{-3.14}^{+3.44}$&    543.8 & 587.2 \\
14&($2.51 $ , $2.60 $)&$0.44_{-0.06}^{+0.06}$&$141_{-27}^{+35}$&$0.91_{-0.09}^{+0.10}$&$662_{-42}^{+49}$&$3.30_{-0.17}^{+0.20}$&$6.95_{-1.64}^{+1.97}$&     553.0 & 596.4 \\
15&($2.60 $ , $2.72 $)&$0.58_{-0.06}^{+0.06}$&$131_{-19}^{+51}$&$1.06_{-0.07}^{+0.11}$&$731_{-51}^{+62}$&$3.24_{-0.18}^{+0.23}$&$9.68_{-2.08}^{+2.70}$&     623.0 & 666.4 \\
16&($2.72 $ , $2.86 $)&$0.62_{-0.06}^{+0.04}$&$257_{-81}^{+62}$&$1.33_{-0.22}^{+0.23}$&$747_{-79}^{+94}$&$3.60_{-0.30}^{+0.42}$&$9.07_{-2.04}^{+1.90}$&    569.4 & 612.9 \\
17&($2.86 $ , $3.05 $)&$0.72_{-0.03}^{+0.03}$&$324_{-123}^{+206}$&$1.05_{-0.22}^{+1.69}$&$689_{-64}^{+428}$&$3.05_{-0.16}^{+0.41}$&$10.82_{-1.58}^{+1.80}$& 579.6 & 623.0 \\
18&($3.05 $ , $3.21 $)&$0.56_{-0.06}^{+0.06}$&$144_{-44}^{+89}$&$0.76_{-0.07}^{+0.09}$&$679_{-40}^{+47}$&$2.57_{-0.07}^{+0.085}$&$5.67_{-1.30}^{+1.59}$&       660.5 & 703.9 \\
19&($3.21 $ , $3.39 $)&$0.47_{-0.06}^{+0.06}$&$141_{-25}^{+51}$&$0.83_{-0.08}^{+0.09}$&$614_{-40}^{+40}$&$2.43_{-0.06}^{+0.074}$&$3.55_{-0.83}^{+1.04}$&      596.8 & 640.3 \\
20&($3.39 $ , $3.58 $)&$0.57_{-0.03}^{+0.03}$&$485_{-114}^{+77}$&$1.69_{-0.44}^{+0.46}$&$965_{-122}^{+262}$&$3.04_{-0.15}^{+0.18}$&$5.49_{-0.87}^{+1.01}$&    613.3 & 656.7 \\
21&($3.58 $ , $3.77 $)&$0.52_{-0.04}^{+0.03}$&$408_{-59}^{+63}$&$1.46_{-0.29}^{+0.33}$&$833_{-100}^{+137}$&$2.82_{-0.12}^{+0.14}$&$3.77_{-0.63}^{+0.73}$&  704.2 & 747.6 \\
22&($3.77 $ , $3.89 $)&$0.28_{-0.07}^{+0.09}$&$205_{-55}^{+165}$&$0.85_{-0.16}^{+0.57}$&$645_{-51}^{+137}$&$2.59_{-0.08}^{+1.87}$&$0.99_{-0.04}^{+0.05}$&  654.7 & 698.1 \\
23&($3.89 $ , $3.99 $)&$0.22_{-0.08}^{+0.07}$&$143_{-20}^{+20}$&$0.95_{-0.10}^{+0.10}$&$594_{-39}^{+42}$&$2.62_{-0.08}^{+0.09}$&$2.12_{-0.61}^{+0.75}$&      648.6 & 692.0 \\
24&($3.99 $ , $4.10 $)&$0.35_{-0.07}^{+0.06}$&$177_{-34}^{+32}$&$1.10_{-0.14}^{+0.13}$&$656_{-49}^{+58}$&$3.18_{-0.16}^{+0.20}$&$3.66_{-0.99}^{+1.10}$&     640.2 & 683.6 \\
25&($4.10 $ , $4.23 $)&$0.50_{-0.05}^{+0.06}$&$196_{-35}^{+100}$&$1.16_{-0.16}^{+0.50}$&$574_{-50}^{+162}$&$3.04_{-0.16}^{+0.24}$&$6.27_{-1.28}^{+1.90}$&      639.9 & 683.3 \\
26&($4.23 $ , $4.37 $)&$0.64_{-0.02}^{+0.03}$&$180_{-22}^{+5}$&$1.66_{-0.04}^{+0.18}$&$489_{-64}^{+24}$&$3.15_{-0.13}^{+0.33}$&$13.14_{-1.92}^{+1.17}$&     558.5 & 601.9 \\
27&($4.37 $ , $4.51 $)&$0.81_{-0.04}^{+0.04}$&$162_{-22}^{+39}$&$1.48_{-0.18}^{+0.29}$&$426_{-52}^{+91}$&$2.93_{-0.21}^{+0.33}$&$25.87_{-4.09}^{+4.99}$&     607.2 & 650.6 \\
28&($4.51 $ , $4.65 $)&$0.90_{-0.03}^{+0.03}$&$322_{-55}^{+56}$&$1.88_{-0.40}^{+0.62}$&$596_{-76}^{+276}$&$3.50_{-0.31}^{+1.11}$&$36.83_{-4.60}^{+5.10}$&    584.6 & 628.1 \\
29&($4.65 $ , $4.79 $)&$0.75_{-0.03}^{+0.02}$&$332_{-42}^{+10}$&$1.94_{-0.17}^{+0.12}$&$882_{-39}^{+54}$&$3.95_{-0.33}^{+0.42}$&$18.06_{-1.97}^{+1.91}$&   616.7 & 660.2 \\

\hline

\end{tabular}
\end{center}
\end{tiny}
\end{table*}


\appendix 
\section{Spectral models}
\label{sec:models}
The Band model is given by:
\begin{equation}\label{eq:Band_function}
  \mathop N\nolimits_{{\rm{B}}}(E)  = \left\{ {\begin{array}{*{20}{c}}
  {K_{B}{{\left( {E/100} \right)}^\alpha }\exp \left( { - E\left( {2 + \alpha } \right)/{E_{{\rm{p}}}}} \right)}&{{\rm{,\ }}E < {E_{\rm{b}}}}\\
  {K_{B}{{\left\{ {\left( {\alpha  - \beta } \right){E_{{\rm{p}}}}/\left[ {100\left( {2 + \alpha } \right)} \right]} \right\}}^{\left( {\alpha  - \beta } \right)}}\exp \left( {\beta  - \alpha } \right){{\left( {E/100} \right)}^\beta }}&{{\rm{,\ }}E \ge {E_{\rm{b}}}}
\end{array}} \right.\ ,
\end{equation}

where $E_b =(\alpha-\beta)E_p/(2+\alpha)$. 
The Band model with a high-energy exponential cutoff at $E_c$ (BC) is given by

\begin{equation}\label{eq:BC}
 {N_{{\rm{BC}}}(E)} = \mathop N\nolimits_{{\rm{B}}}(E)\times exp\left(-\frac{E}{E_c}\right).
\end{equation}

Other models considered in this paper include a \sw{blackbody}\footnote{\url{https://heasarc.gsfc.nasa.gov/xanadu/xspec/manual/node137.html}} (BB), blackbody added to Band (B+BB), a broken power-law model with two breaks (\sw{bkn2pow}\footnote{\url{https://heasarc.gsfc.nasa.gov/xanadu/xspec/manual/node141.html}}), a broken power-law model with one break and a high-energy cutoff (\sw{bknpow}C\footnote{\url{https://heasarc.gsfc.nasa.gov/xanadu/xspec/manual/node142.html}}), and a power-law model with a high-energy exponential cutoff added to \sw{blackbody} (BB+CPL\footnote{\url{https://heasarc.gsfc.nasa.gov/xanadu/xspec/manual/node160.html}}), as given by:

\begin{equation}
\label{eq:band_bb}
  N_{B+BB}(E) = N_{B}(E) + K_{BB}\frac{E^2}{exp(E/kT_{BB})-1}\\
\end{equation}
  
\begin{equation}
\label{eq:bkn2pow_function}
  \mathop N\nolimits_{{\sw{bkn2pow}}}(E)  = \left\{\begin{array}{*{20}{c}}
  {K^b {E}^{-\alpha_{\rm{1}}}} & {{\rm{,\ }}E \le {E_{\rm{1}}}}\\
  {K^b {E_{\rm{1}}}^{\alpha_{\rm{2}} - \alpha_{\rm{1}}} {E}^{-\alpha_{\rm{2}}}} &{{\rm{,\ }}  {E_{\rm{1}} \le E \le {E_{\rm{2}}}}}\\
{K^b {E_{\rm{1}}}^{\alpha_{\rm{2}} - \alpha_{\rm{1}}}{E_{\rm{2}}}^{-\beta - \alpha_{\rm{2}}} {E}^{\beta}} & {{\rm{,\ }E \ge {E_{\rm{3}}}}}\\
\end{array} \right.\ ,
\end{equation}

\begin{equation}
\label{eq:bkn2powC}
  {N_{{\sw{bkn2powC}}}(E)} = \mathop N\nolimits_{{\sw{bkn2pow}}}(E)\times exp\left(-\frac{E}{E_c}\right).
\end{equation}

\begin{equation}
\label{eq:bknpow}
    N_{\sw{bknpow}}(E) = \left\{\begin{array}{*{20}{c}}
    {K {E}^{-\alpha_{\rm{1}}}} & {{\rm{,\ }}E \le {E_{\rm{1}}}}\\
    {K {E}_1^{\alpha_{\rm{2}}-\alpha_{\rm{1}}}}({E}/1keV)^{-\alpha_2} & {{\rm{,\ }}E > {E_{\rm{1}}}}\\
\end{array} \right.\
\end{equation}

\begin{equation}
\label{eq:bknpowC}
    {N_{{\sw{bknpowC}}}(E)} = \mathop N\nolimits_{{\sw{bknpow}}}(E)\times exp\left(-\frac{E}{E_c}\right).
\end{equation}

\begin{equation}
\label{eq:bb_cpl}
    N_{BB+CPL}(E) = K_{BB}\frac{E^2}{exp(E/kT_{BB})-1} + K_C E^{-\xi} exp(-\frac{E}{E_c})\\
\end{equation}

\section{\oldgrb LAT-HE: sub-GeV cutoff and transition from prompt to afterglow} \label{sec:afterglow_transition}
A high-energy cutoff (around 100 $MeV$) to the Band function is found to be the best-fit model in the case of \oldgrb. In some other GRBs also, a sub-GeV cutoff is seen \citep{Abdo:2009ApJ, Tang:2015ApJ, Vianello:2017}. LAT emission can be divided into two major episodes (arising due to prompt and afterglow) and these episodes can superimpose. For \oldgrb, we observe a transition from prompt to afterglows and note some of the properties shown by the emission observed in LAT. 

\begin{figure*}
\centering
\includegraphics[scale=0.5]{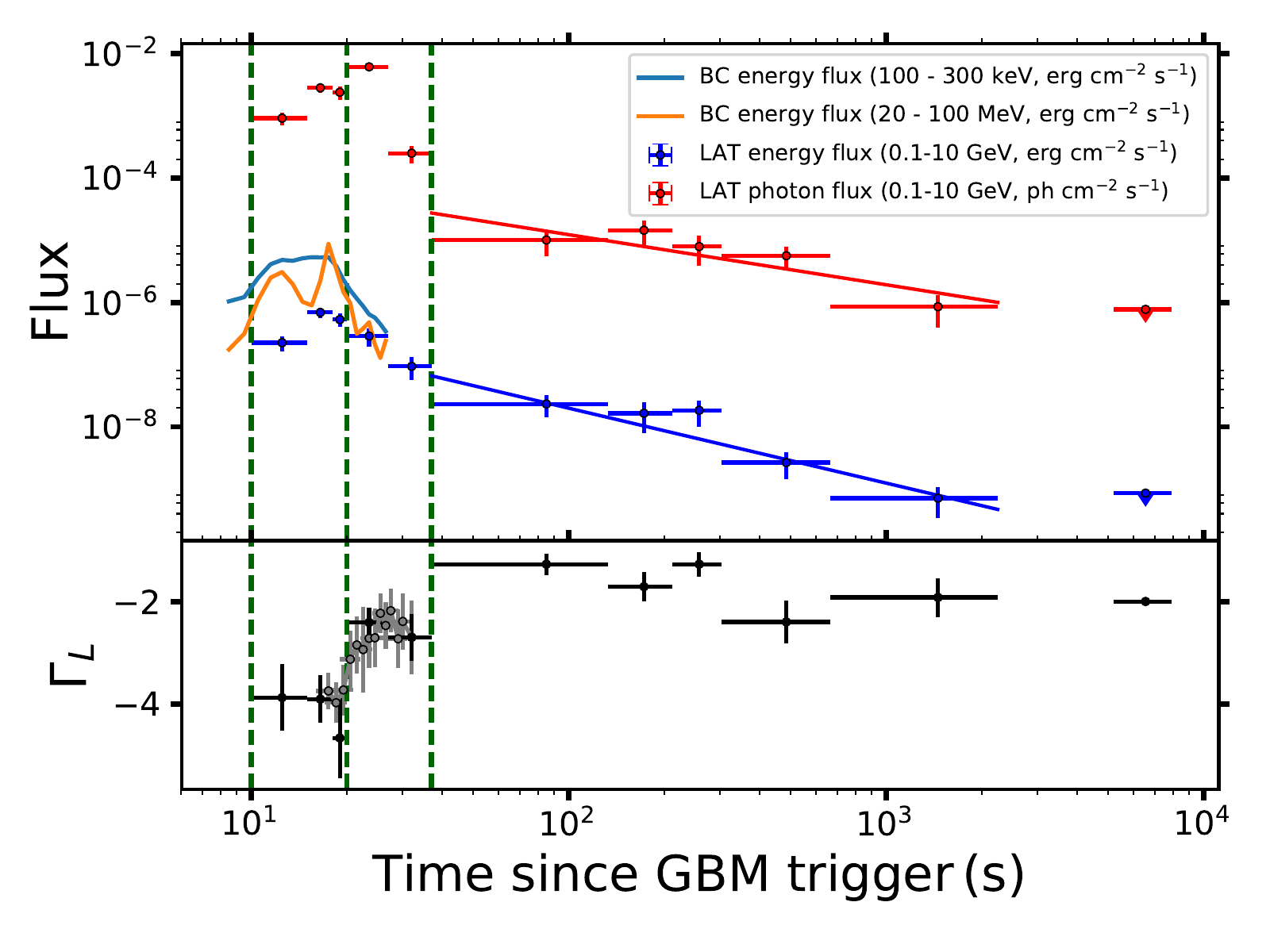}
\includegraphics[scale=0.5]{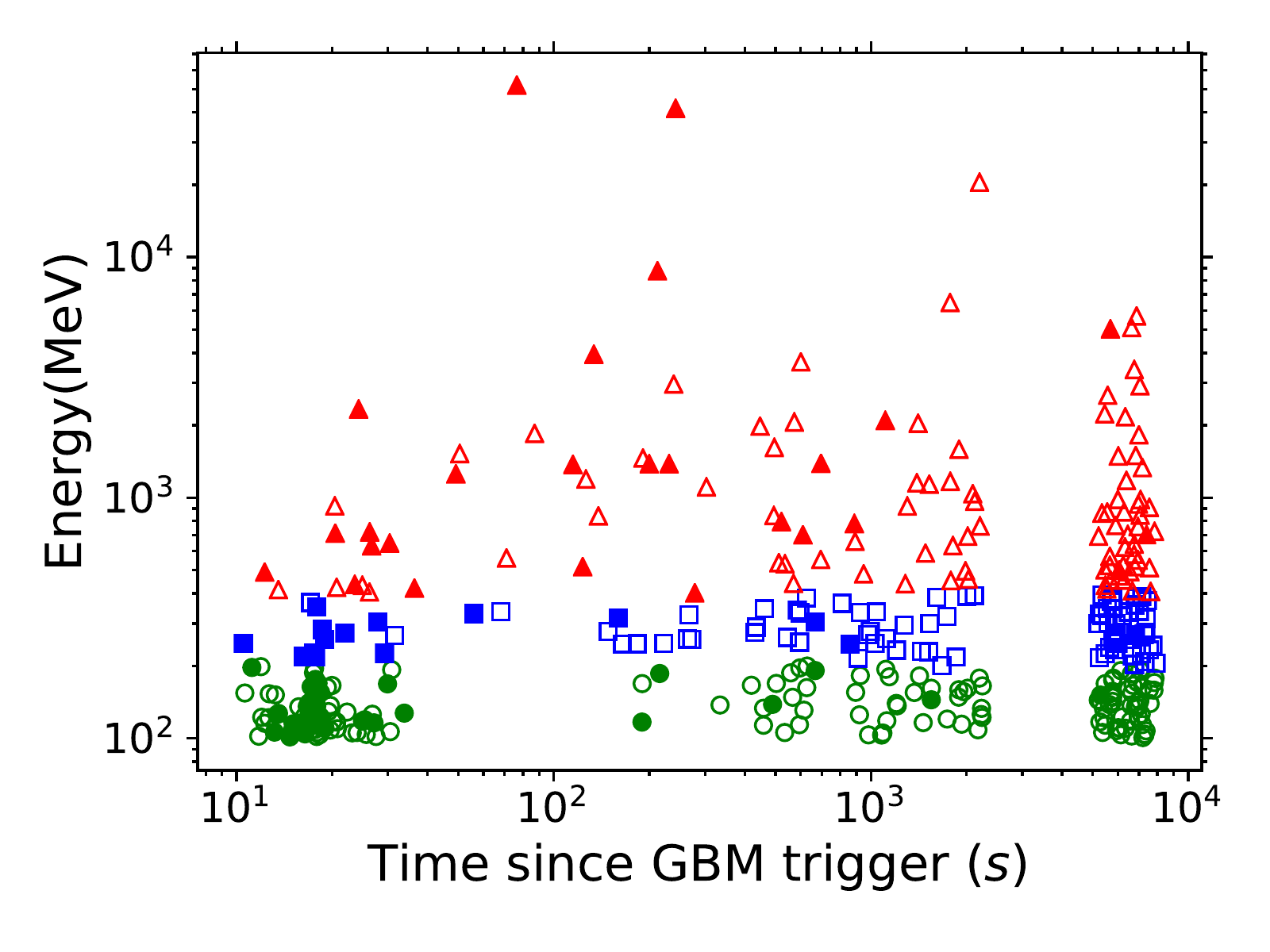}
\caption{\oldgrb: (\emph{Left}), flux evolution in the energy band 0.1 - 10 $GeV$ is shown. There are two distinct components in the LAT ($>$ 100 $MeV$) emission, a fast varying (FV) prompt emission component and a slow varying (SV) afterglow component. The afterglow  component is engraved into the prompt emission and is subdominant. The photon index smoothly changes from softer to harder values showing the transition from the FV to the SV component. \emph{Right} The light curve in three different energy ranges is shown here. The filled data points are the photons that have $\ge 90\%$ probability of association with \oldgrb and the open data-points are ones with probability $<90\%$. Circles denote photons with energy between 100 $MeV$ and 200 $MeV$,  squares represent photons between 200 $MeV$ and 400 $MeV$, and stars represent photons with energy $>$ 400 $MeV$. The first photon with energy greater than 400 $MeV$ (and thus spectrally lying on the harder power law) is at $T0$ + 12.27 $s$. It is coincident with the peak of the first LLE pulse (20-100 $MeV$). The second photon with a greater energy 711 $MeV$ is at 20.5 $s$ and coincident with the end of second LLE pulse. This point in time can be the beginning of a temporal component different from the prompt emission}
\label{fig:lat_flux_evolution}
\end{figure*}

\begin{enumerate}
    \item Initially, the emission is observed simultaneously with GBM. The LAT-HE flux evolution shows two components with different hardness (see the left panel of Fig.~\ref{fig:lat_flux_evolution}), the former being a softer emission that lasts until $\sim 40 s$ and the latter being a harder extended emission. The former is a fast varying (FV) component since its photon flux varies with time as $\propto t^{-3.98 \pm0.53}$, the latter being a slow varying (SV) LAT-HE component that may extend to an earlier time, thus showing us hardening of the spectral component as a result of its superposition with the earlier softer emission.
    \item We, here, track the hard component by monitoring the LAT emission during the overlapping time-window, which helps us smoothly observe the evolution of the spectral index. The FV component is soft and in the time-integrated spectrum can be thought to be the spectrum above the cutoff in the Band spectrum. The two components are dominant in different energy regions. The FV component is majorly populated by the photons with energies that are less than 200 $MeV$ whereas the SV component by ones with energy greater than 200 $MeV$. The light curves in Fig.~\ref{fig:LCs} (right panel) show photons with energies near 1 $GeV$ that are first observed after $\sim$ 20 $s$. This implies the presence of LAT-HE afterglow starting earlier than or beginning from 20 $s$. This claim can further be supported by the flux evolution of LAT in the energy range 0.1-10  $GeV$ as seen in Fig.~\ref{fig:lat_flux_evolution}.
    \item  By looking at the evolution of spectral index, we clearly see the transition from prompt to afterglow emission. In wider bins, this soft to hard transition can be seen in the 20 - 27 $s$ and 27 - 37 $s$ bins. To see this as a smooth transition we made narrower bins of 3 $s$ duration and used the sliding window technique with a step of 1 $s$, or 2 $s$ (for the last few bins). We plot the spectral index of the power-law fit obtained for these windows. The index evolves from a softer value observed in the bins 8  - 13, 13 - 15, and 15 - 18 $s$ to a harder value observed for the bins after 37 $s$ (see Fig~\ref{fig:lat_flux_evolution}).
\end{enumerate}

To summarize, sub-GeV spectral cutoffs may not be very evident in the time-integrated spectrum of the prompt emission. However, if the LAT photon index shows soft to hard evolution, this is a first hint of a spectral cutoff at sub-GeV energies. The time-resolved spectra can show a cutoff in the spectral bins where it is actually present. The transition from the prompt to afterglows can be studied by sliding a coarse bin with size which has a well-constrained photon index. The evolution of the photon index would become significant as the external component will start shaping the spectrum.
The start point can also be marked by observing the photon light curve in an unbinned likelihood analysis of the LAT data.




    


\clearpage
\bibliography{GRB190114C_ref} 
\bibliographystyle{aasjournal}

\end{document}